\documentclass[trackchanges]{aastex701}

\usepackage{comment}
\usepackage{hyperref}
\usepackage{booktabs}
\usepackage{ulem}
\begin{document}

\title{Exploring the long-term temporal variability in polarization through multi-epoch optical spectro-polarimetry - Part I: A sample of Herbig Ae/Be and classical Be stars}

\author[orcid=0000-0001-7340-8873,sname='Maiti']{Arijit Maiti}
\affiliation{Physical Research Laboratory, Ahmedabad, India, 380009}
\affiliation{Indian Institute of Technology Gandhinagar, Gandhinagar, India, 382055}
\email[show]{arijitmaiti@prl.res.in}  

\author[gname=Mudit K.,sname='Srivastava']{Mudit K. Srivastava} 
\affiliation{Physical Research Laboratory, Ahmedabad, India, 380009}
\email[show]{mudit@prl.res.in}

\begin{abstract}

Polarization signatures across emission line features, together with their temporal evolution, offer a powerful probe of the circumstellar environments of astrophysical sources, on spatial scales otherwise inaccessible to direct imaging techniques. However, the photon-hungry nature of spectro-polarimetry has significantly limited the availability of such datasets in the literature. This work reports a multi-epoch spectro-polarimetric monitoring campaign targeting a sample of Herbig Ae/Be and classical Be stars. The initial observations were obtained during the commissioning and performance-verification phase of ProtoPol, a recently developed medium-resolution echelle spectro-polarimeter mounted on the Physical Research Laboratory (PRL) 2.5m telescope, Mt Abu, India. Given the limited number of comparable datasets available for this class of objects, follow-up observations of the same targets were carried out repeatedly over more than 28 months (December 2023–March 2026), enabling an investigation of the temporal behavior of their polarimetric properties. Our sample comprises 11 Herbig Ae/Be stars and 10 classical Be stars. Our observations found that, while the H$\alpha$ polarization remained relatively constant for the classical Be stars, they showed significant variability for most of the Herbig stars in the sample. The observations presented here constitute one of the rare multi-epoch spectro-polarimetric datasets spanning more than two years and should be of considerable interest to the broader astronomical community. This paper is Part-I of a two-part series of sample studies; corresponding results for symbiotic and red giant stars are presented in Part-II.

\end{abstract}

\keywords{\uat{Spectro-polarimetry}{1278} --- \uat{Herbig Ae/Be stars}{1834} --- \uat{Classical Be stars}{142} }


\section{Introduction} 
\label{sec:Introduction}

Spectro-polarimetry has been found to be a potent technique to probe astrophysical situations that have a directional dependence, such as magnetic fields, asymmetries in the scattering environment, etc. Such asymmetries cause observable signatures in their polarization spectra, which can be measured to derive the underlying physical parameters. For example, the presence of strong magnetic fields can lead to Zeeman splitting of the emission/absorption features in the stellar spectra, which show circular polarization signatures \citep{kuzmychov2017first, phan2009magnetic, fouque2023spirou}. Asymmetries in the scattering environment also give rise to linear polarization signatures in particular emission/absorption lines in the spectra \citep{ababakr2016linear}.  Despite their pivotal role in advancing astrophysical research, the spectro-polarimetry technique is not so prevalent and is often considered to be a specialized one, mostly due to its photon-hungry nature, which necessitates the need for long exposures to achieve the desired signal-to-noise ratio (SNR) and the requirements of large aperture telescopes. This is, in particular, true for scenarios requiring higher resolution observations. For example, the determination of magnetic field strengths and direction requires high-resolution ($>$ 60000) spectro-polarimeters with the capability to measure circular polarization, e.g., SPIRou ($\sim$70000) \citep{donati2020spirou}, ESPaDOnS (resolution $\sim$ 68000) \citep{donati2003espadons}, etc., mounted on moderate (3.5m - 4 m) or larger aperture telescopes. However, due to a relatively smaller number of such large-aperture telescope facilities, this also presents a scheduling bottleneck for high-cadence spectro-polarimetric observations or dedicated observational campaigns. 
\par 
Given such a situation, low- to intermediate-resolution spectro-polarimetry ($<$ 15000) is well suited for the smaller aperture telescopes (2m - 3m). It has been found to be the most optimal way for studying the polarization specific to particular broad emission lines caused by various astrophysical phenomena such as scattering \citep{schmid1994raman, oudmaijer2001export}.  Some examples of such spectro-polarimeters, developed for smaller-aperture telescopes (1m - 3m), are VESPolA (Very Precise Echelle SpectroPolarimeter, R $\sim$8000) at the 1.3 m Araki Telescope in the Koyama Astronomical Observatory, Japan \citep{arasaki2015very}, LIPS (LIne Polarimeter and Spectrograph, R $\sim$7000) at the 2.2 m University of Hawaii Telescope \citep{ikeda2003development}, etc. Such medium-resolution linear spectro-polarimeters were found to be of great utility and have produced many excellent science results like spectro-polarimetry of symbiotic stars, novae, Herbig stars, Be stars, etc \citep{ikeda2004polarized, kawakita2019high, oudmaijer1999halpha}. Thus, to exploit the untapped potential of the technique of spectro-polarimetry, in particular in medium resolution on small-aperture telescopes, an instrumentation program to develop spectro-polarimeters for Physical Research Laboratory (PRL), 1.2m and 2.5m telescopes at Mt Abu Observatory, India was initiated \citep{kumar2022designs}. A medium-resolution echelle spectro-polarimeter, named ProtoPol, has been developed for PRL's 1.2m and 2.5m telescopes \citep{srivastava2024development, srivastava2026development, maiti2026development}. ProtoPol is operating with a spectral resolution between 0.4-0.75 $\AA$ and covering the entire visible wavelength range from 4000-9600 $\AA$. The instrument was developed fully in-house with completely off-the-shelf optical components. It served as a precursor to another currently under development echelle spectro-polarimeter for PRL 2.5m telescope called Mt. Abu Faint Object Spectrograph and Camera - Echelle Polarimeter (M-FOSC-EP), operating in the entire visible wavelength range with a higher resolution of $\sim$15000 \citep{kumar2022designs}.
\par
ProtoPol was commissioned on PRL 1.2m telescope in December 2023 and on 2.5m telescope in February 2024. A variety of astrophysical sources were observed with ProtoPol under its commissioning, performance verification, and science verification programs, such as Herbig Ae/Be stars, classical Be stars, symbiotic stars, red giants, etc \citep{maiti2026development}. The sample includes standard polarized and unpolarized stars, as well as scientifically interesting sources that are typically known to show spectro-polarimetric signatures. Though the sample objects selected for the performance verification of ProtoPol were of various types ranging from hot young Herbig stars to cool red giants, they all possess morphologies conducive to the scattering phenomena, thereby generating polarization signatures. Such morphologies present the most suitable conditions to investigate how stars interact with their surrounding environment, and it is directly linked to the system's evolution. While direct imaging of such systems is still challenging, as the relevant scales are often of the order of submilli-arc-seconds \citep{vink2002probing}, spectro-polarimetry bridges this gap by tracing polarization across emission lines to probe otherwise inaccessible inner regions. Temporal variation of the spectro-polarimetric signatures is, thus, of immense importance as they give insight into the dynamic evolution of the scattering environment in such systems. The above is the basis of the observations presented here covering two different sets of objects, namely,  (i) Herbig Ae/Be $\&$ classical Be stars, and (ii) cool red giant stars $\&$ symbiotic systems. While these two sets appear to be very distinct in their physical mechanism, it is the scattering-based polarization observed in their spectra whose temporal variation would provide interesting insights into their evolution. Recently, one such study was carried out for the science verification of ProtoPol, of the symbiotic star Y Gem, where the orbital phase-dependent variation in the H$\alpha$ polarization signature was observed using multi-epoch spectro-polarimetric observations from ProtoPol \citep{maiti2026discovery}. 
\par
Therefore, all the objects observed during the performance verification of ProtoPol were again observed after a period of nearly one to two years to probe the dynamic scattering environments of the stars by measuring the polarization in the most prominent H$\alpha$ emission and in other emission lines, continuum polarization, etc. As it is realized that such a multi-epoch data set is rather rare in the literature, we seek to inform the community of these observations and results in the form of two manuscripts/research articles. The present manuscript-I (hereafter Paper-I) discusses the results obtained for the sample of hot stars like Herbig Ae/Be stars and classical Be stars, whereas the results for the cool stars are presented in a subsequent manuscript-II (hereafter Paper-II). Below we describe the rationale behind the multi-epoch studies of Herbig Ae/Be stars and classical Be stars.
\par
In the present manuscript, a large sample of Herbig and Be stars was observed with ProtoPol during its performance and science verification program. While polarization signatures have been detected and reported in many Herbig stars included in our sample \citep{oudmaijer1999halpha, vink2002probing, vink2005probing, mottram2007difference}, a multi-epoch spectro-polarimetric variability study of the stars has rarely been attempted, such as \cite{harrington2009spectropolarimetric} who studied the H$\alpha$ spectro-polarimetry of several Herbig and other emission line stars by comparing the results obtained using HiViS spectro-polarimeter with archival ESPaDOnS spectro-polarimeter data. Herbig stars are known to have dynamically evolving circumstellar scattering environments, leading to changing H$\alpha$ polarization profiles and strengths, as reported in various spectro-polarimetric observations \citep{oudmaijer1999halpha, vink2002probing, harrington2009spectropolarimetric, alecian2013high}. Similarly, the dedicated multi-epoch spectro-polarimetric observations of classical Be stars have also been scarce in the literature. Given this lacuna, we decided to conduct a second set of spectro-polarimetric observations of the sources we had earlier observed, at an interval of nearly one to two years, to probe the temporal variability, if any, of polarization in their emission features. The aim of this exercise was to produce a list of sources that are bright enough for their spectro-polarimetry with small-aperture telescopes and are variable in nature. As we shall be discussing later, we indeed discovered the variability  - in particular in the polarization of H$\alpha$ emission - in a good number of sources. For a few sources, the observed polarization was presented in \cite{maiti2026development} as part of ProtoPol's performance/science verification. Here we present the complete multi-epoch spectro-polarimetric observation campaign of these stars, observed over a period of over two years, which should be of general interest to the astronomical community.
\par
The sample presented here consists of 11 Herbig Ae/Be stars and 10 classical Be stars, whose spectro-polarimetric observations span 60 nights (from December 2023 to March 2026). The stellar types included in the sample have a wide variety of circumstellar environments, with Thomson scattering the typical cause of H$\alpha$ polarization observed in Herbig and classical Be stars. While comparing with older spectro-polarimetric data of the stars in the sample, several stars show a different polarization profile than what had been reported in the literature, thus indicating the dynamic scattering environments present in these sources. The paper has been arranged in the following way: the observation campaign is detailed in section~\ref{sec:ObservationCampaign}. The results of Herbig Ae/Be stars are presented in section~\ref{sec:Herbig}, with a brief discussion of the spectro-polarimetric properties for these stars. The results of classical Be stars are presented in section~\ref{sec:classicalBe}, with the results compared to their Herbig counterparts. Finally, section~\ref{sec:summary} summarizes the data reported in this paper. 

 %%%%%%%%%%%%%%%%%%%%%%%%%%%%%%%%%%%%%%%%%%%%%%%%%%%%%%%%
 %%%%%%%%%%%%%%%%%%%%%%%%%%%%%%%%%%%%%%%%%%%%%%%%%%%%%%%%
 
\section{Observations and Data Reduction} 
\label{sec:ObservationCampaign}

Optical spectro-polarimetric observations were conducted using ProtoPol, a medium-resolution echelle spectro-polarimeter currently mounted on the PRL 2.5m telescope at the Mt. Abu Observatory, Gurushikhar, India \citep{kumar2022designs, srivastava2024development, srivastava2026development, maiti2026development}. ProtoPol provides continuous spectral coverage across the visible wavelength range of 4000–9600 $\AA$ with a spectral resolution in the range 0.4–0.75 $\AA$. The instrument integrates a polarimetric module with a spectrometer module employing echelle and cross-disperser (CD) gratings to generate cross-dispersed spectra of the ordinary (o) and extraordinary (e) rays for each echelle order on a $1K \times 1K$ ANDOR CCD detector. Two distinct CD gratings are utilized: one optimized for the blue spectral region (4000–6200 $\AA$) and the other for the red spectral region (5800–9600 $\AA$). The system further incorporates a dedicated calibration unit containing a Uranium–Argon (UAr) lamp for wavelength calibration of the echelle orders and a halogen lamp for order tracing. A complete spectro-polarimetric observation sequence consists of science exposures acquired at four half-wave plate (HWP) orientations: $0^\circ$, $22.5^\circ$, $45^\circ$, and $67.5^\circ$, with calibration frames collected following each exposure. The raw data acquired with ProtoPol are reduced using a fully automated, in-house-developed data-reduction pipeline \citep{maiti2026development}, built with custom Python routines based on open-source libraries like NumPy, SciPy, Astropy, etc. The reduction procedure includes bias and dark subtraction, cosmic-ray rejection, order tracing, subtraction of scattered background and sky contributions, extraction of o- and e-ray intensities for individual orders, wavelength calibration, and derivation of the Stokes parameters from the extracted orthogonally polarized spectra. The details of the data-reduction process and pipeline are described in detail in \cite{maiti2026development}.
\par
The observation campaign was carried out over a period of 28 months (December 2023 - March 2026) since the first light of the instrument, and the first set of observations was conducted in parallel to the instrument's characterization and performance verification observations. A large sample of Herbig Ae/Be stars and classical Be stars was observed over two epochs to understand the multi-epoch spectro-polarimetric variability of the sources across the H$\alpha$ emission feature. A total of 11 Herbig stars and 10 classical Be stars were targeted in the sample study. The criteria for sample selection were the apparent magnitudes of the stars ($V$ $\lesssim$ $11$  for observations from 2.5m telescope and $V$ $\lesssim$ $7$ for observations from 1.2m telescope), the presence of H$\alpha$ emission in their optical spectra, and their relative position in the sky during the observational campaign.  The data for some of the stars (first set) in the sample have been reported in \cite{maiti2026development} and have been reused in this manuscript as well for the sake of continuity. The observation logs for Herbig stars and classical Be stars are presented in Tables~\ref{HerbigObservationLog} and \ref{classicalBeObservationLog}, respectively.

%%%%%%%%%%%%%%%%%%%%%%%%%%%%%%%%%%%%%%%%%%%%%%%%%%%%%

\begin{table*}
\caption{Observation log of Herbig Ae/Be Star sample. All V-magnitudes and spectral types are adopted from SIMBAD\footnote{\href{https://simbad.u-strasbg.fr/simbad/sim-fbasic}{https://simbad.u-strasbg.fr/simbad/sim-fbasic}}.} \label{HerbigObservationLog}
\centering
\setlength{\tabcolsep}{5pt}
\renewcommand{\arraystretch}{1.2}
	\begin{tabular}{cc cc cc cc cc cc}
        \hline
		\hline

\textbf{Name} & \textbf{HD number} & \textbf{Vmag} & \textbf{Spectral} & \textbf{Date} & \textbf{Exposure} \\
 &  &  & \textbf{type} & \textbf{(dd-mm-yyyy)} & \textbf{sec $\times$ HWP pos.} \\
\hline
\hline
27 CMa & 56014 & 4.65 & B4Ve\_sh C & 02-12-2024 & 900s x 4 x 2 sets \\
       &       &     &             & 20-03-2026 & 900s x 4 x 1 set \\

MWC 442 & 13867 & 7.71 & B5IIIe C & 01-12-2024 & 900s x 4 x 3 sets \\
        &       &      &          & 17-01-2026 & 900s x 4 x 1 set \\

AB Aur  & 31293  & 7.05  &  A0Ve C  & 28-11-2024 & 900s x 4 x 4 sets\\
        &        &      &           & 19-12-2025 & 900s x 4 x 1 set \\
        
MWC 480 & 31648 & 7.62 & A5Vep C & 30-11-2024 & 900s x 4 x 5 sets \\
        &       &      &         & 16-01-2026 & 900s x 4 x 1 set \\
        
MWC 758 & 36112 & 8.27 & A8Ve C  & 02-12-2024 & 900s x 4 x 4 sets \\
        &       &     &          & 17-12-2025 & 1800s x 4 x 1 set \\

MWC 120 & 37806 & 7.90 & B9/9.5II/III E & 29-11-2024 & 900s x 4 x 4 sets \\
        &       &     &          & 19-12-2025 & 900s x 4 x 1 set \\

FS CMa & 45677 & 8.50 & B2IV/V[e] C & 29-12-2024 & 600s x 4 x 1 sets \\
        &       &     &             & 20-01-2026 & 900s x 4 x 1 set \\

MWC 158 & 50138 & 6.67 & A1Ib/II C & 29-11-2024 & 900s x 4 x 1 set \\
        &       &     &            & 19-12-2025 & 480s x 4 x 1 set \\
        
GU CMa & 52721 & 6.59 & B2Vne C & 02-12-2024 & 600s x 4 x 2 sets \\
        &       &     &            & 20-01-2026 & 1200s x 4 x 1 set \\
        
HD 58647 & 58647 & 6.85 &  B9IV C & 28-12-2024 & 1800s x 4 x 1 set \\
         &        &     &           & 20-03-2026 & 1800s x 4 x 1 set \\

MWC 147 & 259431  & 8.72 & 	B6ep D   & 28-12-2024 & 1800s x 4 x 2 sets \\
         &        &      &           & 18-02-2026 & 1800s x 4 x 1 set \\

%          \hline
%    \end{tabular}
% %    \caption{Caption}
% %    \label{tab:my_label}
% \end{center}
% \end{table}
% \end{landscape}
		\hline
		\hline
	\end{tabular}
\end{table*}
%\end{landscape}
%\end{deluxetable*}
%\end{rotatetable}
%%%%%%%%%%%%%%%%%%%%%%%%%%%%%%%%%%%%%%%%%%%%%%%%%%%%%%%%%%%%%%%

%%%%%%%%%%%%%%%%%%%%%%%%%%%%%%%%%%%%%%%%%%%%%%%%%%%%%%%%%%%%%%%

\begin{table*} 
\caption{Observation log of classical Be Star sample. All V-magnitudes and spectral types are adopted \href{https://simbad.u-strasbg.fr/simbad/sim-fbasic}{SIMBAD}. The exposure times of the first epoch of observation in most of the targets are typically longer that of the second epoch, as those observations were conducted when ProtoPol was mounted on PRL 1.2m telescope during the early on-sky characterization phase of the instrument.} \label{classicalBeObservationLog}
\centering
\setlength{\tabcolsep}{5pt}
\renewcommand{\arraystretch}{1.2}
	\begin{tabular}{cc cc cc cc cc cc}
        \hline
		\hline

\textbf{Name} & \textbf{HD number} & \textbf{Vmag} & \textbf{Spectral} & \textbf{Date} & \textbf{Exposure} & \textbf{Depolarization} \\
 &  &  & \textbf{type} & \textbf{(dd-mm-yyyy)} & \textbf{sec $\times$ HWP pos.} & \\
\hline
\hline
C Per & 25940 & 4.03 & B3Ve C      & 18-12-2025 & 50s x 4 x 1 set & No\\
       &       &     &             & 11-04-2026 & 180s x 4 x 1 set & No \\

$\beta$ CMi & 58715 & 2.89 & B8Ve C & 05-01-2024 & 300s x 4 x 1 set & No \\
            &       &      &        & 17-02-2026 & 120s x 4 x 1 set & No \\

$\gamma$ Cas  & 5394  & 2.39  &  B0.5IVpe C  & 31-12-2023 & 300s x 4 x 1 set & Yes\\
              &        &      &              & 17-02-2026 & 60s x 4 x 1 set & Yes \\
        
$\epsilon$ Aur & 31964 & 2.99 & A9Ia C & 05-01-2024 & 300s x 4 x 2 sets & No \\
        &       &      &               & 17-02-2026 & 120s x 4 x 1 set & No \\
        
$\beta$ Mon & 45725 & 4.60 & B4Veshell C & 16-01-2024 & 600s x 4 x 2 sets & Yes \\
        &       &     &                  & 10-04-2026 & 300s x 4 x 1 set  & Yes \\

$\omega$ Ori & 37490 & 4.59 & B3Ve C     & 18-12-2025 & 180s x 4 x 1 sets & No \\
        &       &     &                  & 11-04-2026 & 480s x 4 x 1 set & No \\

$\phi$ Per  & 10516 & 4.06 & B1.5V:e-shell C & 24-01-2024 & 300s x 4 x 2 sets & Yes \\
        &       &     &                      & 17-02-2026 & 150s x 4 x 1 set & Yes \\
        
$\psi$ Per & 22192 & 4.23  & B5Ve C & 24-01-2024 & 300s x 4 x 2 sets  & Yes\\
        &       &     &             & 17-02-2026 & 150s x 4 x 1 set  & Yes\\
        
$\zeta$ Tau & 37202 & 3.03 &  B1IVe$\_$shell C & 12-01-2024 & 120s x 4 x 3 sets  & Yes \\
         &        &     &                   & 17-02-2026 & 120s x 4 x 1 set &  Yes \\
  
$\eta$ Tau  &  23630  & 2.87 & 	B7III C & 25-01-2024 & 180s x 4 x 2 sets  & No \\
         &        &      &             & 18-03-2026 & 90s x 4 x 1 set & No \\

%          \hline
%    \end{tabular}
% %    \caption{Caption}
% %    \label{tab:my_label}
% \end{center}
% \end{table}
% \end{landscape}
		\hline
		\hline
	\end{tabular}
\end{table*}
%\end{landscape}
%\end{deluxetable*}
%\end{rotatetable}
%%%%%%%%%%%%%%%%%%%%%%%%%%%%%%%%%%%%%%%%%%%%%%%%%%%%%%%%%%%%%%%

\par
However, the typical polarization variation measured across the emission line features is of the order of a fraction of a percent \citep{oudmaijer1999halpha, vink2005probing}. Therefore, for statistically significant deduction, the polarization errors should be kept small, typically of the order of  $\sim$0.1-0.3$\%$. As the instrumental polarization of ProtoPol is established to be less than $0.1\%$ \citep{maiti2026development}, it is, thus, a suitable instrument for such studies. The errors in degree and angle of polarization ($\sigma_P$ and $\sigma_{\theta}$) are dependent on the signal-to-noise-ratio (SNR) of each spectral resolution element \citep{patat2006error}, and is given by the equations: $\sigma_P = \frac{1}{\sqrt{N/2}(S/N)}$ and $\sigma_{\theta} = \frac{\sigma_P}{2p}$ where N is the number of HWP positions in which the star has been observed for Stokes parameter determination and (S/N) is the required SNR. For ProtoPol, with N = 4, the typical polarization errors of $\sim$0.1-0.3$\%$ demand the SNR to be in the range of $\sim$230-700 for each spectral resolution element.
\par 
Moreover, the signal from the peak of the emission feature to the wings would also vary, with high SNR and achieved at the peak, leading to accurate polarization determination as compared to the wing regions or absorption troughs (in the case of P-Cygni profiles or double-peaked line structures, etc.), which suffer from low SNR \citep{oudmaijer1999halpha, vink2002probing}. Thus, the ProtoPol data-reduction pipeline employs a dynamic binning algorithm, where the size of the spectral bin is dynamically increased to add more flux elements until the required SNR for that spectral bin is achieved. This ensures the polarization error corresponding to each spectral bin is constant, at the cost of degradation of spectral resolution. 
\par
The data presented in this paper have not been corrected for instrumental polarization, as it would only have a polarization bias effect on the data. For the same reason, the presented data have also not been corrected for interstellar polarization (ISP), as any variation in ISP with wavelength manifests only over a very broad wavelength range \citep{serkowski1974many}, and as such, can be considered constant across any emission/absorption feature.

%%%%%%%%%%%%%%%%%%%%%%%%%%%%%%%%%%%%%%%%%%%%%%%%%%%%%%%%%%%%%%%%

\section{Herbig Ae/Be stars}
\label{sec:Herbig}

Herbig Ae/Be stars, first identified by \cite{herbig1960spectra}, are intermediate-mass pre-main-sequence stars of spectral types A or B, associated with star-forming regions. They are the most massive pre-main-sequence stars that are observable in the optical and infrared wavelengths and bridge our understanding between the formation of lower-mass solar-type stars ($M_\star$ $\lesssim$ 1.5$M_\odot$) and the most massive stars ($M_\star$ $\gtrsim$ 10$M_\odot$). They exhibit hydrogen Balmer line emission (especially H$\alpha$), nebulosity, and an infrared excess from circumstellar dust \citep{BrittainHerbig}. \cite{de1994new} expanded the original definition proposed by \cite{herbig1960spectra} to include extinction laws, IR excess, photometric variability, and various emission features. Their spectra also show forbidden [O I] emission indicative of winds or outflows, and Fe II multiplets \citep{hernandez2004spectral}. Most Herbig stars form in binaries via disc fragmentation \citep{wheelwright2010mass}, and exhibit spectroscopic variability due to rotational modulation of accretion regions \citep{garcia2016investigating}.
\par 
The process of low-mass star formation by cloud collapse and disc accretion is a relatively well-understood problem in astrophysics. However, whether high-mass stars form similarly remains debatable. Attempts have been made to explore the formation of such stars through disc accretion methods \citep{behrend2001formation}, while \cite{bonnell1998formation} explored the possibility of star formation through stellar collisions and dense cluster environments. To address these questions, intermediate-mass Herbig Ae/Be stars form the bridge between low-mass and high-mass star formation processes, as they are the only higher-mass pre-main-sequence stars that are visible in optical and infrared (IR) wavelengths \citep{vink2002probing}. As such, they form suitable laboratories to explore the formation of high-mass stars, which are otherwise shrouded in the dust envelope and, as such, not visible in optical wavelengths.
\par
Herbig Ae/Be stars, like their classical Be star counterparts, often show linear spectro-polarimetric signatures across the H$\alpha$ and other Balmer emission lines, due to Thomson scattering of emission radiation in the circumstellar environment. Such sources are surrounded by shells, envelopes, disks of material, etc., and their circumstellar material often participates in accretion processes, polar outflows, wind structure, and disk-star interplay, and leaves traces of the prevalent physical mechanism in their polarization spectra \citep{oudmaijer1999halpha, vink2002probing}. First established in studies of classical Be stars \citep{clarke1974observations}, spectro-polarimetry revealed that depolarization across the H$\alpha$ profile arises because H$\alpha$ photons are emitted over a more extended volume than continuum photons, resulting in fewer scatterings. The technique has since been applied to B[e] stars, LBVs, supernovae, and novae \citep{zickgraf1989polarization, schulte1994axisymmetric, cropper1988spectropolarimetry, bjorkman1994spectropolarimetry}. For pre-main-sequence stars, \cite{oudmaijer1999halpha} found that roughly half of Herbig Be stars show spectro-polarimetric signatures across the Balmer lines, while \cite{vink2002probing} extended the sample to include Herbig Ae stars and disentangled viewing-angle effects from intrinsic geometry. Subsequent high-resolution studies using ESPaDOnS \citep{harrington2009spectropolarimetric, alecian2013high} and multi-line optical observations \citep{ababakr2016linear} further characterized the polarization profiles of Herbig Ae/Be stars.
\par
The origin of the H$\alpha$ polarization in Herbig Ae stars, like T Tauri stars, results from emission line photons scattered off a rotating accretion disc \citep{vink2005probing, hubrig2006accurate}. The photons originating from the hotspot on the stellar photosphere would be polarized as they scatter off the disc material \citep{vink2005probing}. On the other hand, Herbig Be stars show an entirely different kind of H$\alpha$ polarization signature, with the continuum more polarized compared to the emission line. This kind of `depolarization' signature, common in classical Be stars, is consistent with the idea of an extended ionized emission region, thus suffering fewer scatterings as compared to continuum photons, and therefore, shows a decrease in polarization \citep{oudmaijer1999halpha}. Thus, in summary, emission line photons are intrinsically polarized in Herbig Ae stars compared to the continuum, whereas in Herbig Be stars, a depolarization signature relative to continuum polarization is expected. The line effect for intrinsically polarized H$\alpha$ profiles is usually narrower than the emission feature, as compared to depolarization signatures, which are as wide as the emission feature itself \citep{ababakr2016linear} and are typically associated with a loop in the $(q, u)$ plane. 
\par
Apart from that, some of the stars in the sample do not show any change in polarization across the H$\alpha$ emission feature. This is caused by a circularly symmetric projection of the circumstellar environment of the star in the sky plane, either due to a spherically symmetric distribution of circumstellar matter itself, or if the disc is viewed pole-on \citep{vink2002probing}. Another special effect is seen in some Herbig stars, which show a strong emission feature with a blue-shifted absorption component. Known as McLean effect \citep{mclean1979interpretation}, a stronger polarization is observed across the absorption trough as compared to the continuum or the emission part. This happens because a residual flux persists in absorption features due to an isotropic re-emission process, which scatters some photons back into the line of sight. In an aspherical scattering medium, this results in an enhanced polarization in the absorption component \citep{ababakr2016linear}. 
\par
The observation log of the observed sample of Herbig Ae/Be stars is given in Table~\ref{HerbigObservationLog}. The derived spectroscopic and spectro-polarimetric line parameters like equivalent width (EW), line peak to continuum ratio, degree and angle of polarization of the continuum, and presence/absence of any noticeable line effect are presented in Table~\ref{HerbigParameters}. The degree and angle of polarization of the continuum were estimated by taking the median value of the neighboring region $\sim$ 20$\AA$ away from the line center of the H$\alpha$ emission feature. The continuum polarization in Herbig Ae/Be stars is caused by the scattering of stellar photons by polarizing agents like dust and electrons \citep{vink2002probing}. The observed continuum polarization is a cumulative effect of both intrinsic polarization caused by asymmetric circumstellar geometry and that due to interstellar media in the line of sight of the observer. Moreover, continuum polarization variability is also a commonplace phenomenon and has been noted by \cite{vrba1979observations}. The same is found in our sample as well for several sources from multi-epoch observations. The multi-epoch spectro-polarimetric data across the H$\alpha$ emission of the sample are shown in Figure~\ref{Fig-Herbig1}, \ref{Fig-Herbig2}, and \ref{Fig-Herbig3}. All the corresponding \textit{q-u} maps across the H$\alpha$ emission are presented in Figure~\ref{Fig-Herbig_qu_1} and \ref{Fig-Herbig_qu_2} in Appendix~\ref{Appendix:qu_maps}.

%=====================================
\begin{table*} 
\caption{Herbig Ae/Be stars derived spectroscopic and spectro-polarimetric parameters} \label{HerbigParameters}
\centering
\setlength{\tabcolsep}{5pt}
\renewcommand{\arraystretch}{1.2}
	\begin{tabular}{c ccccc ccccc}
           \hline
		  \hline
          
        %\toprule
        \multicolumn{1}{c}{\textbf{Name}} & \multicolumn{5}{c}{\textbf{Epoch-1}} & \multicolumn{5}{c}{\textbf{Epoch-2}} \\
        \cmidrule(lr){2-6} \cmidrule(lr){7-11}
        & H$\alpha$ EW & Line peak/cont. & $p_{cont} $ & $\theta_{cont}$ & Line & H$\alpha$ EW& Line peak/cont. & $p_{cont} $ & $\theta_{cont}$ & Line \\
        &  ($\AA$) & ratio & $(\%)$ & (deg) & effect? &  $\AA$ & ratio & $(\%)$ & (deg) &  effect? \\
        %\midrule
        \hline
		  \hline
        27 CMa & -5.140 $\pm$ 0.250 & 1.429 $\pm$ 0.023 & 0.66 & 32.6 & Yes & -7.149 $\pm$ 0.397 & 1.468 $\pm$ 0.044 & 1.17 & 75.5 & No \\

        MWC 442 & -8.476 $\pm$ 0.059 & 3.702 $\pm$ 0.023 & 1.34 & 53.4 & No & -10.750 $\pm$ 0.088 & 4.015 $\pm$ 0.033 & 1.22 & 57.4 & No \\
        
        AB Aur & -27.239 $\pm$ 0.171 & 5.737 $\pm$ 0.033 & 0.73 & 42.7 & No & -21.629 $\pm$ 0.089 & 3.713 $\pm$ 0.020 & 0.39 & 26.3 & Yes \\
        
        MWC 480 & -21.892 $\pm$ 0.079 & 4.778 $\pm$ 0.017 & 0.45 & 16.6 & Yes & -22.458 $\pm$ 0.154 & 5.391 $\pm$ 0.028 & 0.56 & 4.9 & Yes \\
        
        MWC 758 & -21.429 $\pm$ 0.158 & 5.143 $\pm$ 0.028 & 1.00 & -10.9 & Yes & -16.965 $\pm$ 0.105 & 3.946 $\pm$ 0.019 & 0.38 & -46.6 & Yes \\

        MWC 120 & -32.082 $\pm$ 0.251 & 6.551 $\pm$ 0.041 & 0.47 & 28.9 & Yes & -27.349 $\pm$ 0.191 & 5.625 $\pm$ 0.032 & 0.59 & 62.9 & Yes \\

        FS CMa & -128.718 $\pm$ 2.877 & 20.451 $\pm$ 0.391 & 0.71 & -64.1 & Yes & -144.508 $\pm$ 3.222 & 27.741 $\pm$ 0.537 & 0.57 & -36.4 & Yes \\

        MWC 158 & -62.460 $\pm$ 0.811 & 14.391 $\pm$ 0.134 & 0.81 & -70.8 & Yes & -68.204 $\pm$ 0.864 & 13.995 $\pm$ 0.136 & 0.55 & -70.3 & Yes \\

        GU CMa & -13.556 $\pm$ 0.246 & 2.773 $\pm$ 0.031 & 1.21 & -26.9 & No & -12.041 $\pm$ 0.277 & 2.603 $\pm$ 0.035 & 1.27 & -18.4 & No \\

        HD 58647 & -11.699 $\pm$ 0.069 & 2.862 $\pm$ 0.038 & 0.49 & -71.7 & No & -11.840 $\pm$ 0.248 & 2.768 $\pm$ 0.063 & 0.93 & 31.4 & Yes \\

        MWC 147 & -67.894 $\pm$ 0.930 & 11.813 $\pm$ 0.146 & 1.03 & 41.5 & Yes & -62.594 $\pm$ 1.086 & 11.194 $\pm$ 0.148 & 1.00 & 58.4 & No \\

\hline
\hline
	\end{tabular}

\end{table*}

%=====================================

\subsection{Comments on individual stars}

\begin{figure*}
  \centering
  \includegraphics[width=\textwidth]{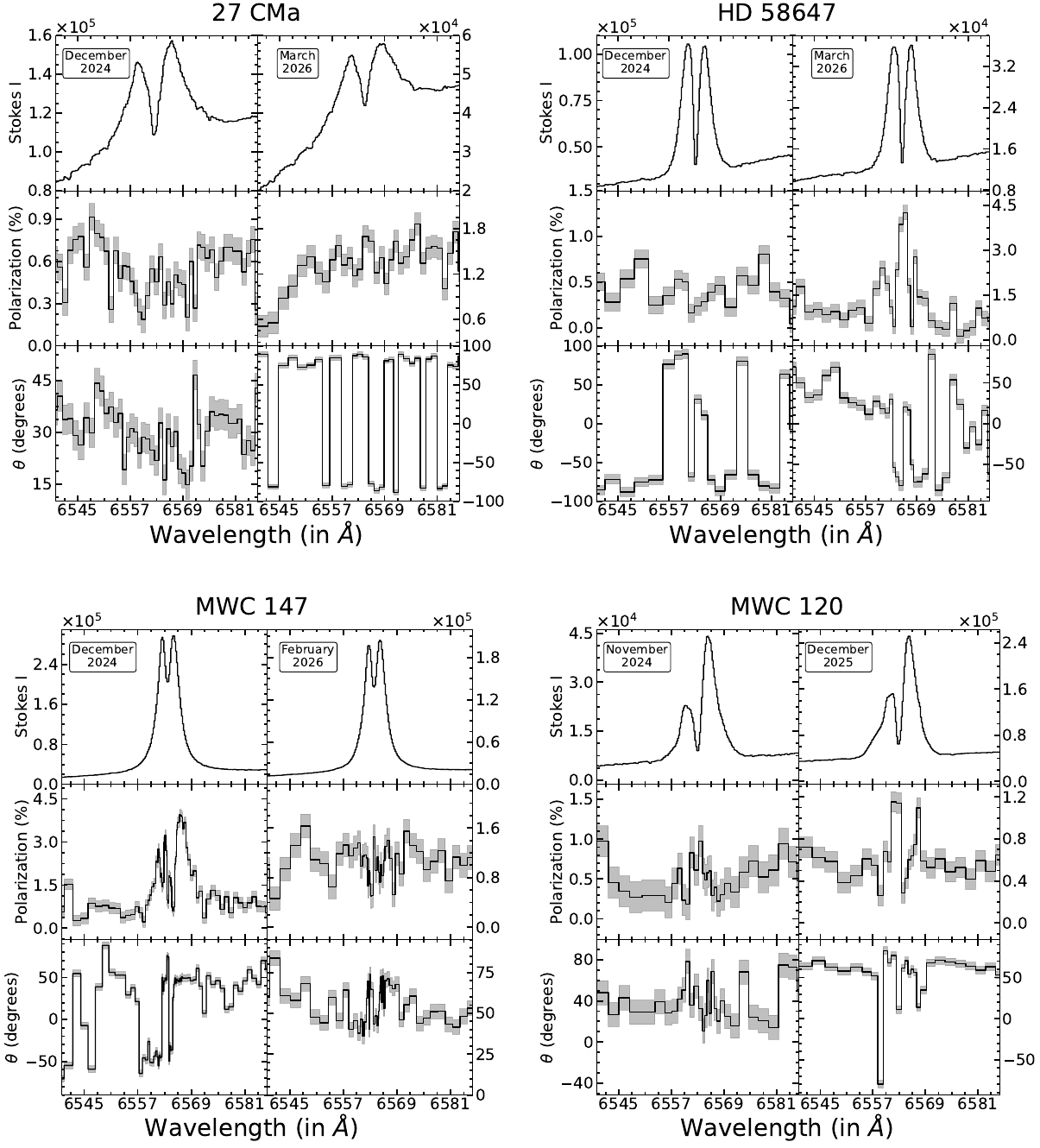}  
  \caption{Polarization profiles for H$\alpha$ as recorded in the spectra of Herbig stars 27 CMa, HD 58647, MWC 147, and MWC 120 over two epochs separated by $\sim$ 12-16 months. Stokes parameter I (total intensity), the degree of polarization, and the angle of polarization are shown in the top, middle, and bottom panels, respectively. The data have been dynamically binned to achieve a constant polarization accuracy over each bin. The corresponding errors in the detection of p and $\theta$ are shown with a gray shaded strip.}
  \label{Fig-Herbig1}
\end{figure*}

\textbf{27 CMa:} It is a known $\beta$-Cep pulsator \citep{balona1991appearance} with a possible binary companion. It is noted to be a shell star with a strongly variable disc and asymmetric spectral evolution \citep{labadie2022classifying}. Observations from ProtoPol (see Figure~\ref{Fig-Herbig1}) show a small but broad depolarization feature across the entire H$\alpha$ profile during the first epoch of observation. However, during the second epoch of observation, no depolarization signature was observed; the polarization gradually increased from the bluer to redder wavelengths.   
\par

\textbf{HD58647:} It is a late B-type star, which was claimed to be a binary by \cite{baines2006binarity} from analysis of the H$\alpha$ centroid and equivalent width (EW). However, monitoring of the star with the HiViS spectro-polarimeter, for over two years, \cite{harrington2009spectropolarimetric} found the H$\alpha$ profile was nearly invariant. Multi-wavelength polarization analysis hints the star has a disk-like structure in the ionized gas surrounding the star \citep{oudmaijer2001export}. \cite{vink2002probing} reported a polarization increase of 0.6$\%$ across the central absorption dip. \cite{harrington2009spectropolarimetric} reported a typical increase in polarization $\sim$0.5$\%$ across the central absorption dip, although the polarization in the red- and blue-shifted peaks was the same as the continuum. Observations from ProtoPol (see Figure~\ref{Fig-Herbig1}) on the first epoch showed almost negligible polarization change across the H$\alpha$ feature. However, observations of the star in the second epoch showed a strong increase in polarization across the absorption component, with peak polarization around 4.5$\%$. The blue and red peaks also show an increase in polarization (peak $\sim$ 3$\%$) with respect to the continuum level.  
\par
\textbf{MWC 147:} A small depolarization ($\sim$0.2$\%$) was reported by \cite{oudmaijer1999halpha} for the star with a continuum polarization of $\sim$1.06$\%$. \cite{vink2002probing} reported a decrease in polarization across the blue peak, but did not classify the effect as depolarization, which is typically associated with B-type stars. \cite{mottram2007difference} also reported spectro-polarimetric observations of the star, albeit with a significantly degraded resolution. Results of multi-epoch spectro-polarimetric observations revealed no polarization detection to small amplitude ($\sim$0.2$\%$) changes across both peaks \citep{harrington2009spectropolarimetric}. Observations from ProtoPol (see Figure~\ref{Fig-Herbig1}), on the other hand, revealed a strong polarization increase 2-3$\%$ above continuum for the first observation epoch, with a possible polarization angle flip across blue-to-red peak. On the other hand, spectro-polarimetric observations in the second epoch, more than 1-year apart, reveal no polarization detection. This shows MWC 147 to be a highly variable H$\alpha$ polarization source, and a possible long-term spectro-polarimetric follow-up of the source is required to understand its changing polarization geometries fully.
\par
\textbf{MWC 120:} The strongly variable H$\alpha$ intensity and polarization profiles were first documented by \cite{oudmaijer1999halpha}, who noted the changing blue/red flux ratio of the emission feature and a clear change in the polarization angle across the central absorption. \cite{vink2002probing} reported a 0.4$\%$ increase in polarization across the red peak above continuum, while \cite{mottram2007difference} reported a 0.2$\%$ decrease in polarization with respect to (w.r.t) continuum across the red peak. \cite{harrington2009spectropolarimetric} presented extensive datasets of the star, including observations from both ESPaDOnS and HiViS spectro-polarimeters \citep{harrington2009spectropolarimetric}, reporting strong polarization changes ($\sim$ 0.5-1.5$\%$ in \textit{q} and \textit{u}) across the central absorption feature. Observations from ProtoPol also confirm the strongly variable polarization nature of the source (see Figure~\ref{Fig-Herbig1}). While the first epoch observations reveal a small ($\sim$ 2.5-3$\sigma$ above continuum) change across the central absorption, the effect is more pronounced ($\geq$ 5$\sigma$ above continuum) in the second epoch. Another distinct polarization component is also visible across the red peak of the H$\alpha$ profile, with a similar amplitude to that observed for the central dip. 

%%%%%%%%%%%%%%%%%%%%%%%%%%%%%%%%%%%%%%%%%%%%%%%%%
%%%%%%%%%%%%%%%%%%%%%%%%%%%%%%%%%%%%%%%%%%%%%%%%%%

\begin{figure*}
  \centering
  \includegraphics[width=\textwidth]{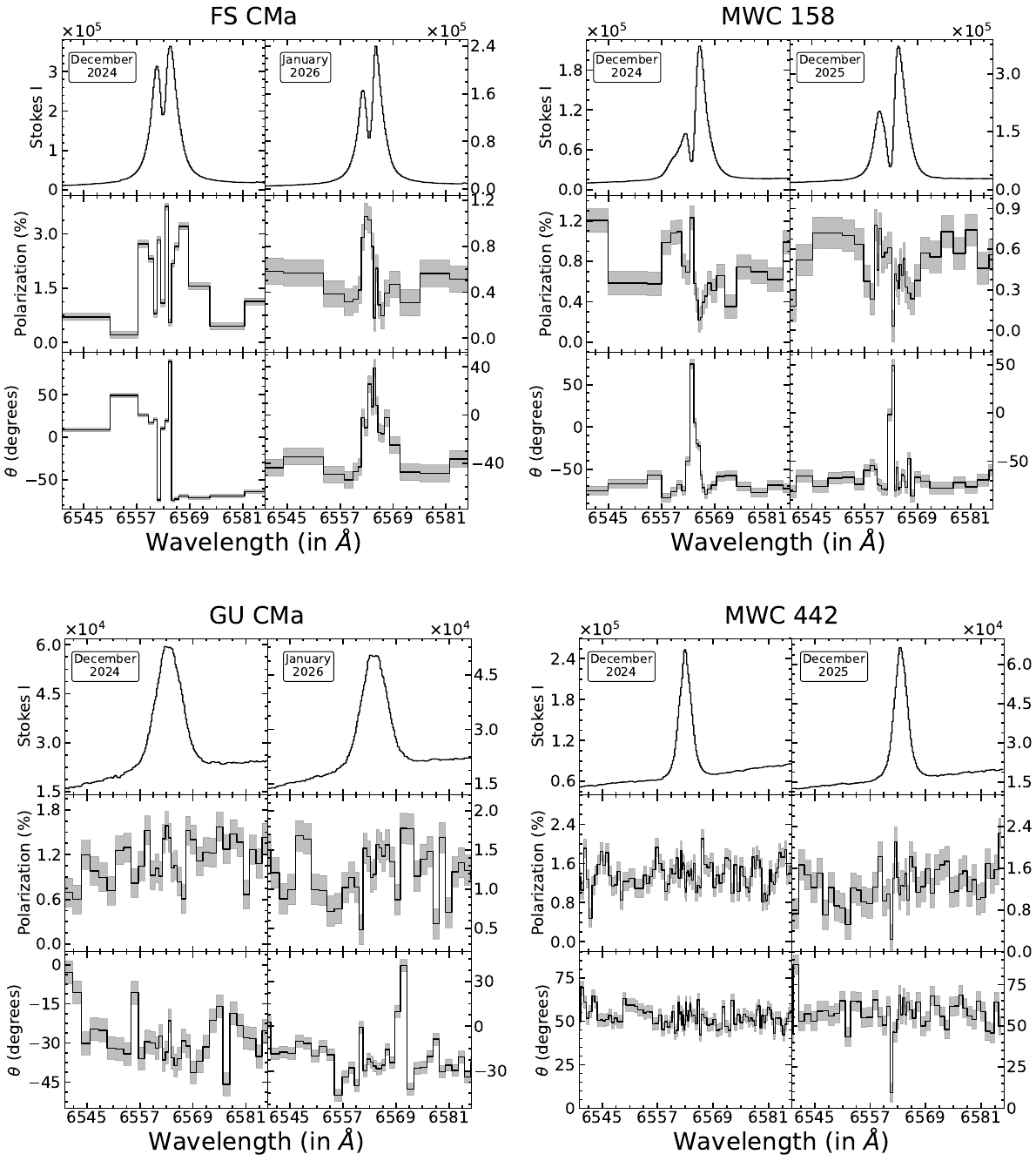}  
  \caption{Same as \ref{Fig-Herbig1}, but for Herbig stars FS CMa, MWC 158, GU CMa, and MWC 442.}
  \label{Fig-Herbig2}
\end{figure*}

\textbf{FS CMa:} While observations of the star by \cite{oudmaijer1999halpha} showed a single-peaked asymmetric H$\alpha$ profile with an instrumental resolution of 40-60 km/s, later observations by \cite{harrington2009spectropolarimetric} showed a double-peaked profile, and the same is observed with ProtoPol. \cite{baines2006binarity} claimed the star to be a binary from their spectro-astrometric results. Spectro-polarimetric observations of the star reported by \cite{oudmaijer1999halpha}, taken over two epochs separated by almost 2 years, showed clear intrinsic polarization signatures along with a change in the angle of polarization across the H$\alpha$ profile. A rotation of the polarization angle 11 to 140 degrees. They predicted circumstellar dust scattering rather than electron scattering as the cause of the variable polarization. A clear polarization signature ($\sim$1$\%$) at the central absorption was also reported by \cite{harrington2009spectropolarimetric} over several epochs of observation. Our observations with ProtoPol (see Figure~\ref{Fig-Herbig2}) also reveal a similar variable polarization trend. Polarization signatures are evident in both epochs of observations. While a strong ($\sim$3$\%$ above continuum) and broad polarization is found across the entire H$\alpha$ profile during the first epoch of observation, narrower intrinsic polarization is noticed for the second epoch, mainly across the central absorption dip. Furthermore, a near 90-degree polarization angle flip is noticed across the profile in the first epoch. In the second epoch, the flip is absent; however, a near 90-degree change in polarization angle is still present across the emission feature. This demonstrates the time evolution of the different polarization components coming into effect in the stellar environment. 

\par
\textbf{MWC 158:} It is a mid-B type star with very strong H$\alpha$ emission (peak intensities $\sim$15 times the continuum flux). \cite{pogodin1997circumstellar} using multi-epoch low-resolution spectroscopic and spectro-polarimetric data found evidence of winds and in-falling matter from analysis of envelope lines. From the near wavelength-independent continuum polarization observed in the star, \cite{bjorkman1998first} claimed electron scattering, rather than dust scattering, being the dominant mechanism for polarization. \cite{baines2006binarity} claimed the star to be a binary from changes in the centroid, and the EW of the H$\alpha$ profile. Spectro-polarimetry of the star was conducted by \cite{oudmaijer1999halpha} over two different epochs, where he reported a depolarization effect across the red peak with a depolarization amplitude of 0.4$\%$ and a continuum polarization of $\sim$0.7$\%$. Observations from \cite{vink2002probing} reveal a small increase in polarization ($\sim$0.3-0.4$\%$ above continuum) across the blue-shifted peak and a depolarization across the red peak. Multi-epoch observations with the HiViS spectro-polarimeter \citep{harrington2009spectropolarimetric} reveal a strong signature ($\sim$1.5$\%$ above continuum for both \textit{q} and \textit{u}) across the central absorption dip, while a broader, weaker component was also discovered across the red peak. Observations with ProtoPol (see Figure~\ref{Fig-Herbig2}) during the first epoch reveal a polarization pattern similar to what was reported by \cite{vink2002probing}, with a small increase in polarization ($\sim$0.6$\%$ above continuum) noticed across the blue peak and a small depolarization ($\sim$0.3-0.4$\%$ below continuum) noticed for the red peak. A similar pattern is also seen in the second epoch of observation, although the magnitude of the effects is smaller than in the earlier epoch. A sharp change in the angle of polarization across the central absorption dip is also seen for both epochs, and a similar effect is reported by \cite{vink2002probing} as well.
\par
\textbf{GU CMa:} The star has been confirmed as a non-detection by both \cite{oudmaijer1999halpha} and \cite{harrington2009spectropolarimetric}, although significant (1.15$\%$) continuum polarization is still present. The star shows an infrared continuum excess \citep{hillenbrand1992herbig}, from which \cite{oudmaijer1999halpha} deduced the star to be of low inclination behind a significant interstellar column. ProtoPol observations (see Figure~\ref{Fig-Herbig2}) validate this claim further, as the star does not show any noticeable polarization change across the single-peaked H$\alpha$ profile for either of the epochs of observation. The determined continuum polarization was in the range 1.0-1.2$\%$ over the two observational epochs.  
\par
\textbf{MWC 442:} It is another B-type Herbig star where no polarization across the single-peaked H$\alpha$ profile was detected \citep{harrington2009spectropolarimetric}. Multi-epoch observations with ProtoPol (see Figure~\ref{Fig-Herbig2}) confirm these results, as the star did not show any polarization change across the H$\alpha$ profile, for both epochs of observation. A continuum polarization of $\sim$1.3-1.4$\%$ was detected and remained stable during both epochs.

%%%%%%%%%%%%%%%%%%%%%%%%%%%%%%%%%%%%%%%%%%%%%%%%%
%%%%%%%%%%%%%%%%%%%%%%%%%%%%%%%%%%%%%%%%%%%%%%%%%%

\begin{figure*}
  \centering
  \includegraphics[width=\textwidth]{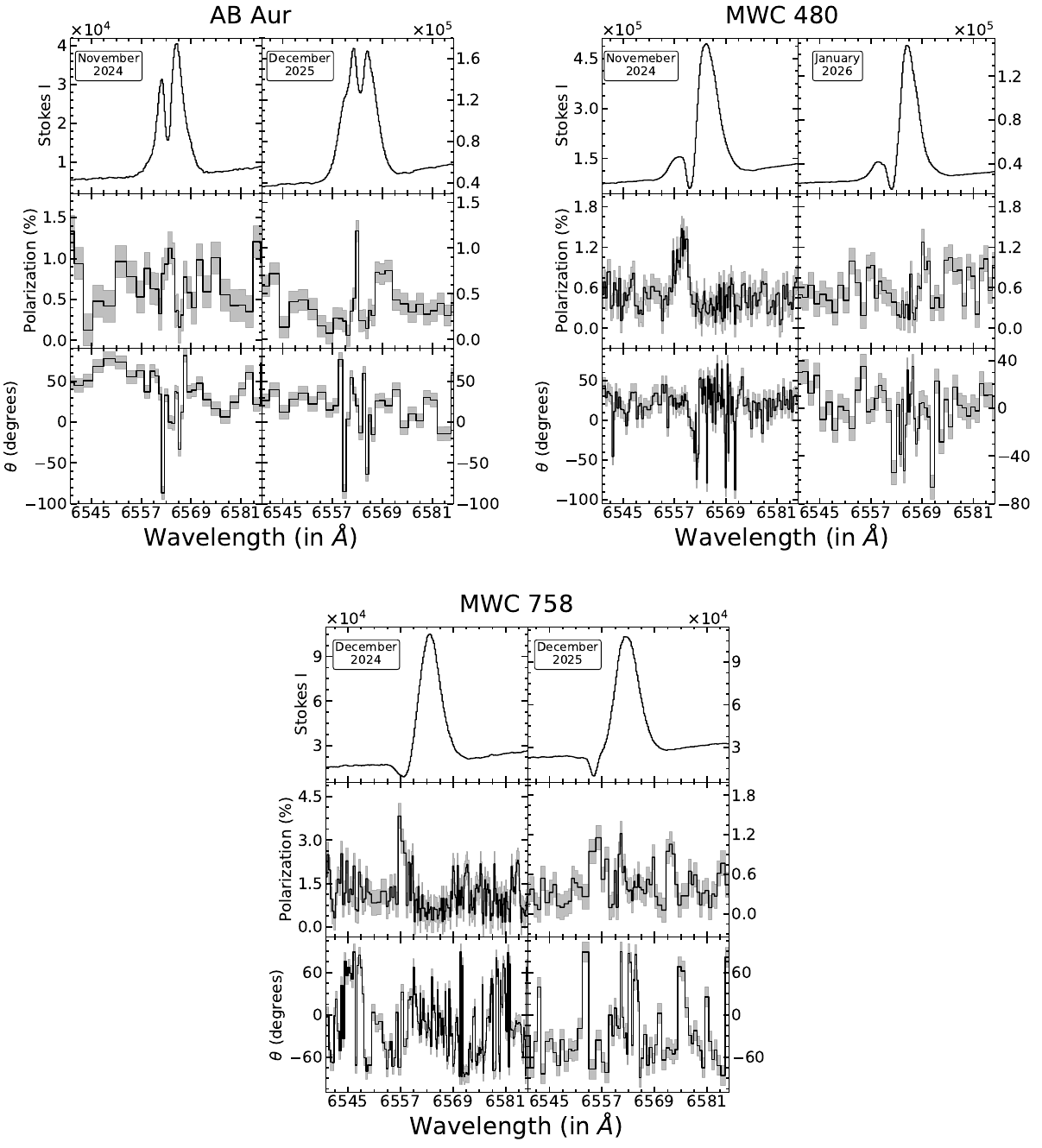}  
  \caption{Same as \ref{Fig-Herbig1}, but for Herbig stars AB Aur, MWC 480, and MWC 758.}
  \label{Fig-Herbig3}
\end{figure*}

\textbf{AB Aur:} It is one of the most well-studied intermediate-mass, late B to early A spectral type Herbig stars, known to show strong polarization in the blue-shifted P-Cygni-like absorption component, an effect known as McLean effect \citep{vink2005probing}. The star has a nearly face-on (inclination angle $\sim$0 degree) circumstellar disc resolved in several wavelengths \citep{grady2005coronagraphic}. Observations of the star by \cite{mottram2007difference} also reported polarization in the absorption component, although the results are somewhat different from those reported by \cite{vink2005probing}. \cite{harrington2009spectropolarimetric} reported 166 spectro-polarimetric observations with HiViS, showing a clear 1$\%$ detection in \textit{q} and \textit{u}. However, high-resolution spectro-polarimetric observations from ESPaDOnS revealed a small 0.2$\%$ level detection with a more complex polarization structure. The author concluded that the source is highly variable in its spectro-polarimetric signatures. Observations with ProtoPol (see Figure~\ref{Fig-Herbig3}) show a markedly different intensity profile than that reported in the above-mentioned literature. Instead of a strong re-shifted peak and a P-Cygni-like blue-shifted absorption component, in the first epoch of observation, we find a more symmetrical double-peaked profile, with a stronger red peak as compared to blue. In the second epoch, the profile becomes more of an equi-strength double-peaked profile; however, a broader emission peak is observed in the bluer side of the blue peak. No clear polarization is detected during the first epoch of observation, while a polarization increase is detected across the central absorption dip in the second epoch, although the increased polarization is sampled by a single resolution element. A small increase is also noticed in the red wing of the H$\alpha$ profile. 
\par
\textbf{MWC 480:} The star has been reported to show clear polarization detection across the blue-shifted absorption component, the so-called McLean effect \citep{ababakr2016linear}. The star has a circumstellar disc, inclined at an angle of 30 degrees \citep{mannings1997rotating}. \cite{beskrovnaya2004active} concluded the presence of an inhomogeneous, azimuthally symmetric wind, variable on time-scales of hours, analyzing H$\alpha$ spectroscopic and continuum polarization data. \cite{vink2002probing} reported a $\sim$0.9$\%$ polarization increase in the absorption trough and a $\sim$0.3$\%$ depolarization across the emission feature above the 0.4$\%$ continuum polarization. Later, in \cite{vink2005probing}, a polarization of $\sim$0.8$\%$ was observed, with continuum polarization varying from 0.18$\%$ to 0.3$\%$. Furthermore, \cite{mottram2007difference} showed a polarization increase of 0.4$\%$ on a 0.2$\%$ continuum polarization. \cite{harrington2009spectropolarimetric} also found a large 1$\%$ polarization detection across the absorption component, with the change in Stokes \textit{u} much wider than that in Stokes \textit{q}. Multi-epoch spectro-polarimetric observations from ESPaDOnS over a period of $\sim$6 months show strong variability in the polarization spectrum, although almost no change is observed in the intensity spectrum. Observations with ProtoPol (see Figure~\ref{Fig-Herbig3}) are in line with earlier conclusions about the polarization variability of the star. A clear polarization $\sim$1.5$\%$ was detected in the first epoch across the blue-shifted absorption component, with continuum polarization $\sim$0.5$\%$. However, almost no polarization was detected across the absorption component in the second epoch, with a slight hint of depolarization observed across the emission part of the profile; the continuum polarization remained nearly constant over the two epochs. 
\par
\textbf{MWC 758:} \citep{beskrovnaya1999spectroscopic} concluded the star to be an unreddened A8V star from multi-color photometry and spectroscopic data, with age $\sim$ 5Myr. They also presented evidence of a gaseous dusty envelope with a variable stellar wind and an extended acceleration zone. Spectro-polarimetric observations with HiViS showed a 1$\%$ change in \textit{u}, with a smaller change in \textit{q} across the blue-shifted absorption component. However, archival data of the star on observations with ESPaDOnS classified it as a non-detection, although \cite{harrington2009spectropolarimetric} notes that the ESPaDOnS observation had significantly less absorption dip than almost all HiViS observations. The star was also a non-detection for the spectro-polarimetric observations reported by \cite{vink2002probing}. A very significant nightly variability $\sim$0.4$\%$ was also observed for the star \citep{beskrovnaya1999ma}. Observations with ProtoPol (see Figure~\ref{Fig-Herbig3}) during the first epoch detected a very strong polarization component $\sim$4$\%$ across the blue-shifted absorption component, thus confirming the detection of McLean effect \citep{ababakr2016linear} in the star. However, in the second epoch, the observed polarization across the blue-shifted absorption is significantly weaker with a peak polarization $\sim$1.2$\%$.

%%%%%%%%%%%%%%%%%%%%%%%%%%%%%%%%%%%%%%%%%%%%%%%%%
%%%%%%%%%%%%%%%%%%%%%%%%%%%%%%%%%%%%%%%%%%%%%%%%%%

\subsection{Summary of Herbig Ae/Be stars multi-epoch spectro-polarimetry}

The multi-epoch spectro-polarimetric observations truly bring out the dynamical nature of the sources. Several of the sources show strong polarization variability across the H$\alpha$ emission feature, even when their intensity spectrum does not show much change. The observed spectro-polarimetric signatures show very different polarization morphologies, even for sources with similar intensity spectra. From our observation sample, 10 out of the 11 Herbig stars showed detectable spectro-polarimetric signatures, at least for one epoch of observation. In many of the stars in the sample, the strongest observed polarization change is noticed across the absorption component in the H$\alpha$ emission line, while the polarization across the emission parts of the line is almost at the level of the continuum polarization. 
\par
Most of the sources showed strong polarization variability over a time span of 12-16 months, except for GU CMa and MWC 442, which did not show any polarization signature across the H$\alpha$ emission line in both the epochs. 27 CMa is also one of the sources that showed a negligible polarization change, although it can be argued that a slight depolarization was detected in the first epoch of observation of the star. AB Aur is one of the stars that surprisingly showed very little polarization signature or change, even though the star has been reported to show strong polarization across its blue-shifted absorption trough. However, from our observations, the line profile of the star has changed significantly since the older observations, with H$\alpha$ emission showing a double-peaked profile instead of a P-Cygni one. Even in between the two observational epochs, the line profile changed significantly, indicating the source to be strongly variable, both in intensity and polarization space. In many of the `windy' systems, which are systems where the H$\alpha$ emission originated from a circumstellar wind as opposed to `disky' systems where the line origin is a circumstellar disk \citep{harrington2009spectropolarimetric}, the clearest spectro-polarimetric variability was observed. This includes sources like MWC 120, MWC 480, and MWC 758, where the polarization change is detected across the central absorption/blue-shifted P-Cygni absorption. In `disky' systems like MWC 158, a difference in the polarization amplitude is evident between blue and red peaks. The other `disky' systems in the sample, like HD 58647, MWC 147, and FS CMa, also show strong polarization variability, with the strongest polarization detected across the absorption trough, but the blue and red emission peaks also show polarization above the continuum level, resulting in an apparent broad spectro-polarimetric signature across the H$\alpha$ emission line.

%%%%%%%%%%%%%%%%%%%%%%%%%%%%%%%%%%%%%%%%%%%%
%%%%%%%%%%%%%%%%%%%%%%%%%%%%%%%%%%%%%%%%%%%%

\section{Classical Be stars}
\label{sec:classicalBe}

Classical Be stars are the most rapidly rotating near main-sequence B-type stars, spinning close to the critical limit where centrifugal force balances gravity \citep{slettebak1988stars}. Their high rotation speeds make them key test beds for studying rotation-induced instabilities \citep{maeder2000evolution}. They are surrounded by an ionized equatorial, outwardly diffusing gaseous Keplerian disc \citep{rivinius2013classical}, optically thick in H$\alpha$ \citep{oudmaijer1999halpha}, and show strong hydrogen Balmer emission. Studies of their circumstellar environments have produced theoretical models applicable to hot stars with non-spherical gas distributions \citep{ignace1996equatorial}. 
\par
In such stars, continuum polarization is primarily produced by electron scattering in ionized, geometrically thin disks, yielding polarization perpendicular to the disk plane. The analytical framework of \cite{brown1977polarisation} describes optically thin, axially symmetric envelopes, where the net polarization depends on the mean electron optical depth, a shape factor characterizing asymmetry, and the inclination angle, predicting an increase in polarization with inclination angle. More realistic models introduce significant deviations from this ideal model, such as absorptive opacity significantly reducing the polarization \citep{fox1991stellar}, while multiple scattering enhances the polarization, leading to peak polarization at intermediate inclinations $\sim$70-80 degrees rather than edge-on systems \citep{wood1996effect}. Consequently, polarization does not scale linearly with density, as attenuation and scattering impose an upper limit, typically $\leq$2$\%$ for Be disks. The spectral dependence of polarization reflects competing opacity sources. Electron scattering dominates at low densities, producing a nearly wavelength-independent polarized continuum, whereas at higher densities, hydrogen photo-ionization and free-free processes dominate. They scale roughly with density squared, becoming significant at higher densities, and reducing polarized flux through pre-scattering absorption \citep{bjorkman1994effects}. This results in an inverse correlation between polarization and H I opacity. 
\par
A comparison between H$\alpha$ line polarization and the continuum provides a powerful diagnostic of circumstellar geometry, as the two components originate in regions of different spatial extent and therefore probe distinct scattering environments  \citep{clarke1974observations, oudmaijer1999halpha}. The H$\alpha$ emission forms over a relatively extended volume within the ionized envelope and experiences minimal additional scattering. In contrast, the continuum radiation, emitted predominantly by the central star and subsequently traversing the envelope, undergoes significantly additional electron scattering. If the projected geometry of the ionized envelope on the plane of the sky is asymmetric, electron scattering induces a net linear polarization in the continuum, while the H$\alpha$ line remains largely unpolarized. This contrast gives rise to a characteristic decrease in polarization across the spectral line, the so-called “depolarization effect” \citep{oudmaijer1999halpha}. Thus, multi-epoch H$\alpha$ spectro-polarimetry of classical Be stars becomes the ideal diagnostic tool for the changing circumstellar morphology of such systems. 
\par
\begin{figure*}
  \centering
  \includegraphics[width=\textwidth]{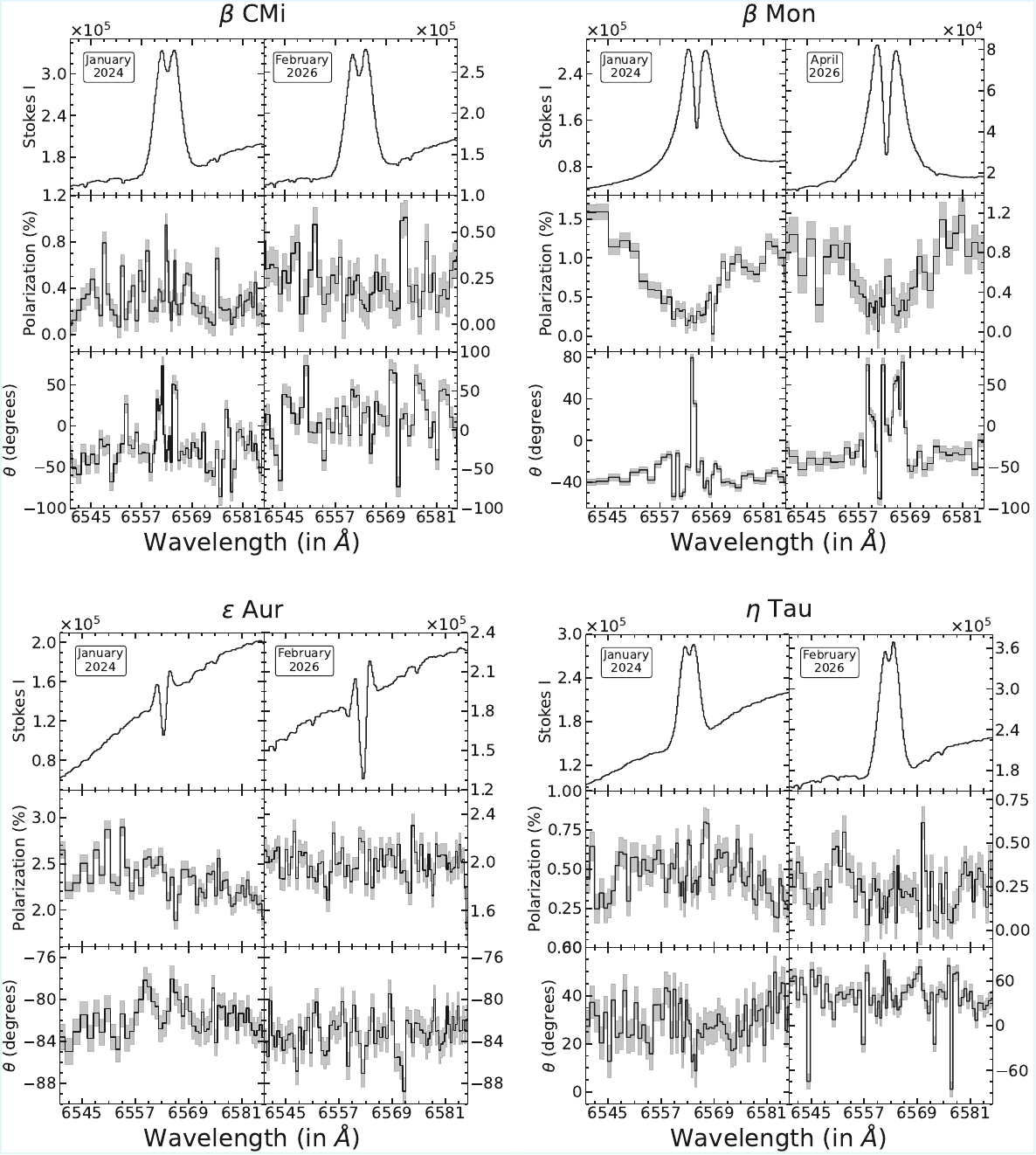}  
  \caption{Same as \ref{Fig-Herbig1}, but for classical Be stars $\beta$ CMi, $\beta$ Mon, $\epsilon$ Aur and $\eta$ Tau. The multiple epochs span over a period of $\sim$ 24-26 months.}
  \label{Fig-Be1}
\end{figure*}
\begin{figure*}
  \centering
  \includegraphics[width=\textwidth]{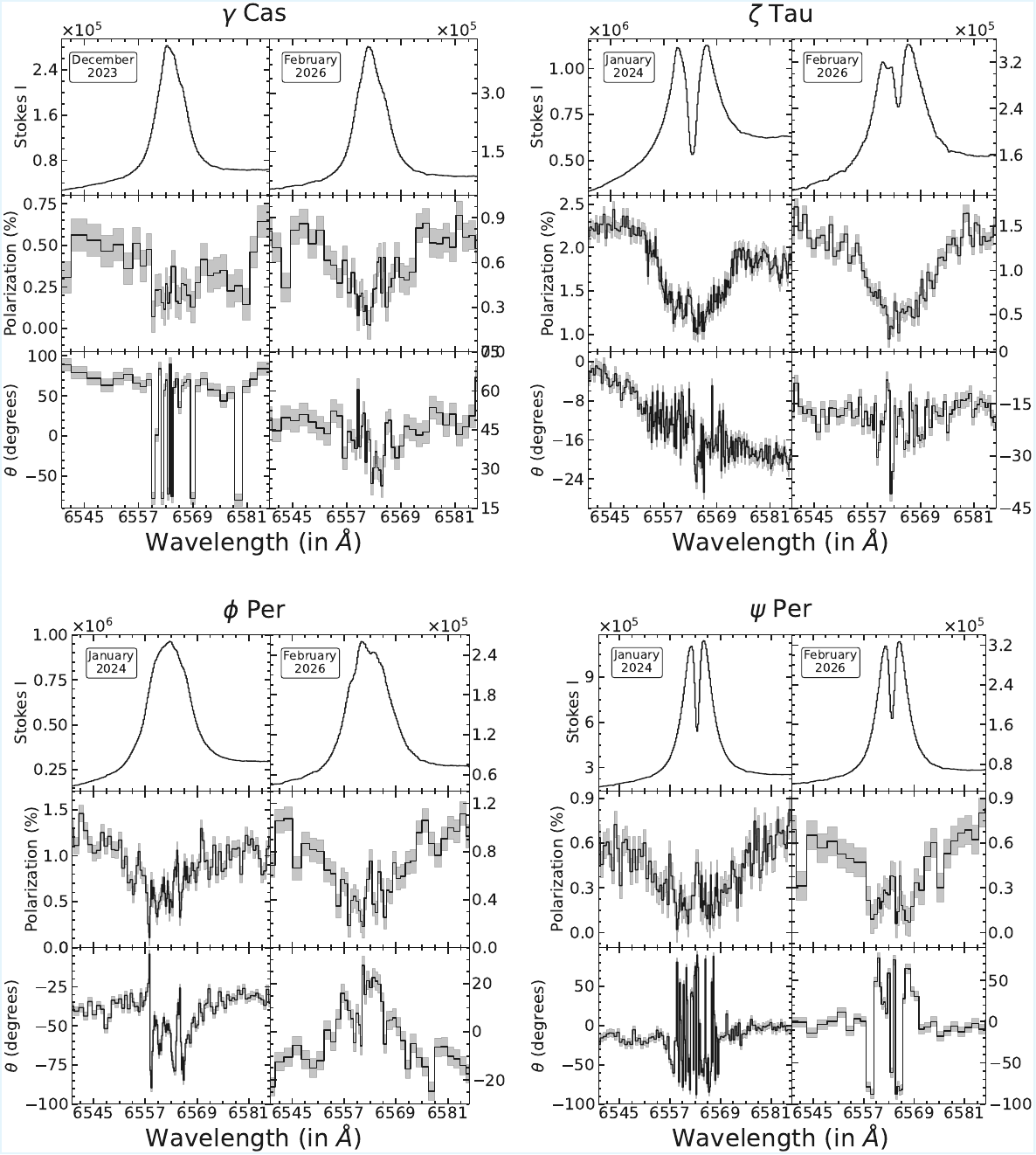}  
  \caption{Same as \ref{Fig-Be1}, but for classical Be stars $\gamma$ Cas, $\zeta$ Tau, $\phi$ Per, and $\psi$ Per.}
  \label{Fig-Be2}
\end{figure*}
\begin{figure*}
  \centering
  \includegraphics[width=\textwidth]{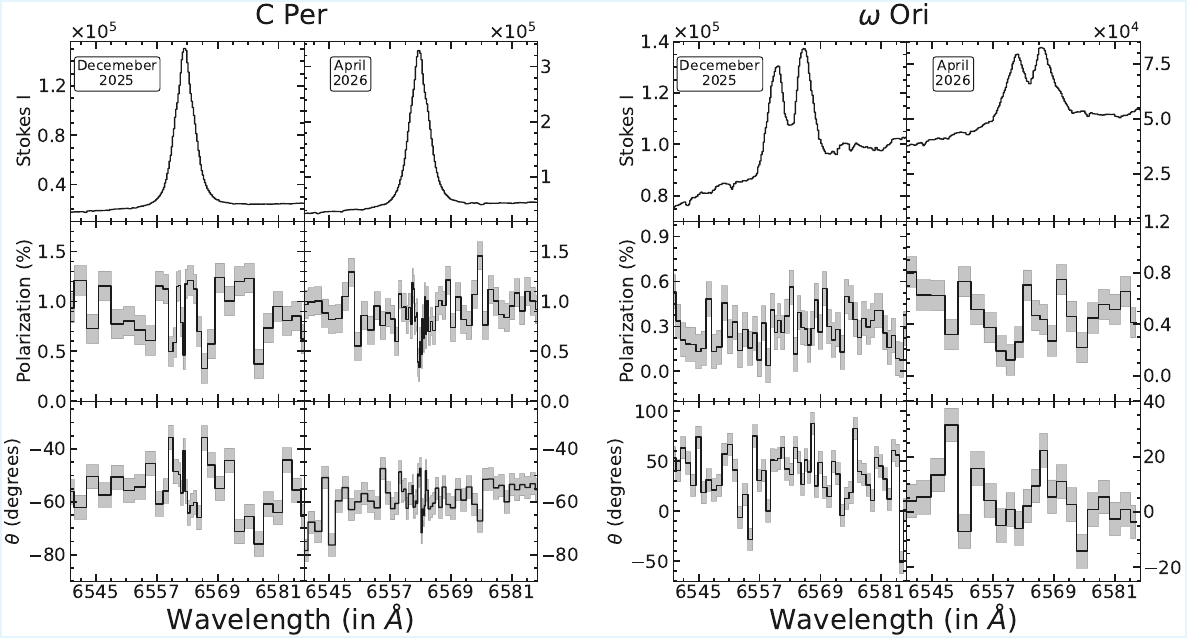}  
  \caption{Same as \ref{Fig-Be1}, but for classical Be stars C Per and $\omega$ Ori.}
  \label{Fig-Be3}
\end{figure*}

The observation log of the observed classical Be stars is given in Table~\ref{classicalBeObservationLog}. The sample was initially chosen for performance verification of the instrument during its characterization phase. Later, a second epoch observation of the targets was also conducted to find the multi-epoch variability of the sources. The multi-epoch  H$\alpha$ spectro-polarimetric observations of the sources are shown in Figures~\ref{Fig-Be1}, \ref{Fig-Be2}, and \ref{Fig-Be3}. The corresponding \textit{q-u} maps across the H$\alpha$ emission are presented in Figures~\ref{Fig-Be_qu_1} and \ref{Fig-Be_qu_2} of Appendix~\ref{Appendix:qu_maps}.

%%%%%%%%%%%%%%%%%%%%%%%%%%%%%%%%%%%%%%%%%%%%
%%%%%%%%%%%%%%%%%%%%%%%%%%%%%%%%%%%%%%%%%%%%

\subsection{Comments on individual stars}

\textbf{$\beta$ CMi:} It is a bright Be star with equi-strength H$\alpha$ double-peaked profile. \cite{quirrenbach1997constraints} reported an extended H$\alpha$ emission region for the source (1.5-3.5 mas), which is much larger than the stellar radius scale. The star shows a very small polarization detection ($\leq$0.1$\%$) with anti-symmetric components, which does not fit with the broad morphology of the typical depolarization signature associated with Be stars \citep{harrington2009spectropolarimetric}. ProtoPol, with a polarization accuracy of $\sim$0.1$\%$, was not able to detect such small polarization signatures. 
\par
\textbf{$\beta$ Mon:} It is a `disky' Be star system that shows a strong and broad depolarization signature across its H$\alpha$, broader than the width of the emission line itself \citep{harrington2009spectropolarimetric}. Unlike its Herbig counterparts, it shows no deviation in polarization trend across the central absorption. Observations with ProtoPol confirm the strong depolarization signature across the H$\alpha$ emission, for both epochs of observation. The depolarization signature remains roughly constant over a period of $\sim$27 months, even though the central absorption becomes deeper during the said period.
\par
\textbf{$\epsilon$ Aur:} It is an eclipsing binary system where the H$\alpha$ line shows a complex emission profile, overlain by strong central absorption which goes below the continuum. Archival data of the source from the ESPaDOnS spectro-polarimeter show strong and complex polarization across the entire width of the absorption component, symmetric about the line center, with $\sim$1.0$\%$ polarization amplitude at the line center \citep{harrington2009spectropolarimetric}. Observations with ProtoPol, however, do not show any noticeable polarization change across the central absorption component for both epochs of observation, even though the star shows strong continuum polarization $\sim$2-2.5$\%$. 
\par
\textbf{$\eta$ Tau:} It is Be star system with an extended H$\alpha$ emission region for the source (1.5-3.5 mas), much larger than the scale of the stellar radii. Like $\beta$ CMi, the star shows a very small ($sim$0.15$\%$) and complex polarization signature, with a small drop in Stokes \textit{u} seen across the line center \citep{harrington2009spectropolarimetric}. Observations with ProtoPol show an almost equi-strength double-peaked H$\alpha$ emission with a small central absorption component. No evident polarization change was detected during the first observation epoch, while a very faint depolarization signature may be noticeable in the second epoch. The continuum polarization varies from $sim$0.25-0.4$\%$.   

%%%%%%%%%%%%%%%%%%%%%%%%%%%%%%%%%%%%%%%%%%%%%%%%%
%%%%%%%%%%%%%%%%%%%%%%%%%%%%%%%%%%%%%%%%%%%%%%%%%%

\textbf{$\gamma$ Cas:} It is a very well-studied Be star system. \cite{quirrenbach1993asymmetric}, using interferometric techniques, resolved the extended H$\alpha$ emission region. The source was heavily monitored by \cite{harrington2009spectropolarimetric} for over a year, where the star did not show any significant H$\alpha$ line profile or spectro-polarimetric variation. It showed a small ($\sim$0.3$\%$) polarization change, and an almost linear excursion in \textit{q-u} space. Observations with ProtoPol showed a broad depolarization across the H$\alpha$ emission line, more clearly in the second epoch as compared to the first. The intrinsic continuum polarization also changed from $\sim$0.5$\%$ to $\sim$0.75$\%$ from the first to the second epoch. This is in line with earlier reported continuum polarization values of the source from 1.0$\%$ \citep{mclean1978polarisation} to 0.8$\%$ \citep{mclean1979interpretation} to 0.5$\%$ \citep{poeckert1977linear}.
\par
\textbf{$\zeta$ Tau:} It is another well-studied Be star system. \cite{quirrenbach1994maximum}, using interferometry techniques, resolved the H$\alpha$ emitting region of the source and constructed a maximum entropy map of the circumstellar region, which showed a highly elongated structure, suggesting a nearly edge-on disc. The star had one of the most variable H$\alpha$ profiles of the sample observed by \cite{harrington2009spectropolarimetric}, with broad spectro-polarimetric effect in both Stokes \textit{q} and \textit{u} parameters of magnitude $\sim$0.5$\%$. Observations with ProtoPol confirm the strong H$\alpha$ line profile variability of the source, with a strong depolarization visible across the emission line for both epochs of observation. The intrinsic continuum polarization of the source also changed from $\sim$2.25$\%$ to $\sim$1.5$\%$ from the first epoch to the second.
\par
\textbf{$\phi$ Per:} It is another classical Be star whose H$\alpha$ emission region is resolved to angular resolutions of 1.5-3.5 mas \citep{quirrenbach1997constraints}. They showed that the emission region is extended as compared to stellar radii, and the same is evident from the strong depolarization signatures across the H$\alpha$ profile of the star for both epochs of observation from ProtoPol. The continuum polarization is $\sim$1.0-1.2$\%$ with polarization decreasing by almost 0.8$\%$ at the line center. The intensity and spectro-polarimetric profiles remain roughly stable between the two epochs. 
\par
\textbf{$\psi$ Per:} It is a `disky' B-type system with an almost equi-strength double-peaked H$\alpha$ profile. \cite{harrington2009spectropolarimetric} reported a hint of an asymmetric spectro-polarimetric signature at the central absorption, with a relative decrease and a relative increase with respect to the broad polarization signature, on the blue and red side of the central absorption, respectively. The author concluded that a combination of linear polarizations, produced from scattering and absorptive components, may be responsible for such polarization morphologies. Observations with ProtoPol, over both epochs of observation, confirm the above results, as an increase in polarization is noticed across the central absorption, overlaid on the broader depolarization signature across the H$\alpha$ emission profile.

%%%%%%%%%%%%%%%%%%%%%%%%%%%%%%%%%%%%%%%%%%%%%%%%%
%%%%%%%%%%%%%%%%%%%%%%%%%%%%%%%%%%%%%%%%%%%%%%%%%%

\textbf{C Per:} It is a B3Ve spectral type Be star with a strong single-peaked emission line. The star does not show any evident polarization signature ($\leq$0.1$\%$) across the H$\alpha$ emission \citep{harrington2009spectropolarimetric}. Observations with ProtoPol confirm this as well, since no evident depolarization was observed across the H$\alpha$ emission for both epochs of observation, where polarization was almost constant with the continuum level at $\sim$1.0$\%$.  
\par
\textbf{$\omega$ Ori:} It is a well-studied `disky' Be star system. \cite{oudmaijer1999halpha} confirmed a non-detection of any spectro-polarimetric signature across its H$\alpha$ profile, with a continuum polarization of 0.3$\%$. The same non-detection was also reported by \cite{vink2002probing} with a continuum polarization of 0.27$\%$. \cite{harrington2009spectropolarimetric} further confirmed it to be a system with no polarization signature detected across the line to an accuracy of less than 0.1$\%$. Observations with ProtoPol also reveal a non-detection for both epochs, with continuum polarization $\sim$0.3$\%$ in the first epoch and slightly higher continuum polarization values ($\sim$0.5$\%$) in the second epoch.

%%%%%%%%%%%%%%%%%%%%%%%%%%%%%%%%%%%%%%%%%%%%%%%%%
%%%%%%%%%%%%%%%%%%%%%%%%%%%%%%%%%%%%%%%%%%%%%%%%%%

\subsection{Summary of classical Be stars multi-epoch spectro-polarimetry}

The multi-epoch spectro-polarimetry of the classical Be star sample, spanning a period of $\sim$28 months, provides a good reference sample to demonstrate the accuracy of the obtained results for the Herbig star spectro-polarimetric sample, due to their more predictable H$\alpha$ polarization behavior. Half of the stars in the sample show depolarization signatures across the H$\alpha$ emission line. The polarization signatures in these stars are less complicated than those in the Herbig sample, either a depolarization signature or no polarization change. Multi-epoch observations further demonstrate that the H$\alpha$ polarization signatures in these stars are a lot more constant as compared to their Herbig counterparts, which showed a far greater temporal variability, both in line profiles and polarization. In a few of the stars, even though there maybe slight changes in the amplitude of the polarization signatures, the overall line effect remained constant. 

\section{Summary} 
\label{sec:summary}

This paper presents the multi-epoch H$\alpha$ spectro-polarimetric observations of a large sample of Herbig Ae/Be stars and classical Be stars, taken over a period of $\sim$28 months. The aim of the study is to present a sample of Herbig Ae/Be and classical Be stars that show (or do not show) any change in their polarization properties across their prominent H$\alpha$ emission over a period of time. We hope that such a sample would be helpful in identifying the sources that could be potential targets of much more extensive studies to explore the physical morphologies of such systems in general. The observations reveal a clear variability of the polarization amplitude and polarization profile across the H$\alpha$ emission for several stars in the sample, especially in the Herbig sample. This includes changes in the intrinsic polarization amplitude or depolarization depth, changes in position angles, and emergence/disappearance of complex line effects. These variations clearly indicate that the scattering geometries and physical conditions within the circumstellar environment are not static, but rather evolve on observable timescales.
\par
For classical Be stars, the variability is broadly consistent, with the targets showing either no polarization signature or a depolarization signature, which remained constant over the observation epochs. On the other hand, the sample of Herbig Ae/Be stars shows diverse spectro-polarimetric effects such as, depolarization, intrinsic polarization, McLean effects, or no changes in polarization across the H$\alpha$, etc. The polarization signatures are also a lot more variable, even on yearly timescales, not just in amplitude but also in polarization profiles, thus demonstrating a lot more complex scattering environments, likely influenced by a combination of gaseous disks, dust, and ongoing accretion or outflow processes.  Thus, the multi-epoch observation of the polarization properties could be useful in deciphering the underlying structure of the circumstellar morphology, and the temporal monitoring is therefore essential for disentangling geometric effects from intrinsic variability. Overall, this study underscores the diagnostic power of H$\alpha$ spectro-polarimetry in identifying the sources with variable circumstellar dynamics and highlights the need for coordinated, long-term time-resolved observations to fully characterize the evolution of such stellar systems.

%% Please use the acknowledgment and contribution environments. This will 
%% be anonomyized when the "anonymous" style option is used. 
\begin{acknowledgments}
The research work at the Physical Research Laboratory (PRL), Ahmedabad, is funded by the Department of Space (DOS), Govt. of India. AM gratefully acknowledges PRL for a Ph.D. research fellowship. PRL operates the Mt. Abu observatory with 1.2 m and 2.5 m telescopes at Mt. Abu. We acknowledge the use of data collected from both the PRL 1.2m and 2.5m telescopes at Mt. Abu Observatory with the ProtoPol instrument. This research has used the SIMBAD database, operated by CDS, Strasbourg, France.
\end{acknowledgments}

\begin{contribution}
AM was responsible for the observation, data reduction, and analysis with ProtoPol. He led the project with initial research concept and manuscript preparation. MKS is the PI of the ProtoPol instrument. He supervised the project and led the project with discussion, manuscript editing, etc. 

\end{contribution}

\facilities{PRL 1.2m and 2.5m telescopes at Mt. Abu Observatory, Gurushikhar, India}



\appendix

\section{Multi-epoch qu plots of the sample of Herbig Ae/Be and classical Be stars}
\label{Appendix:qu_maps}

%%%%%%%%%%%%%%%%%%%%%%%%%%%%%%%%%%%%%%%%%%%%%%%%%
\begin{figure*}
  \centering
  \includegraphics[angle=90, width=\textwidth]{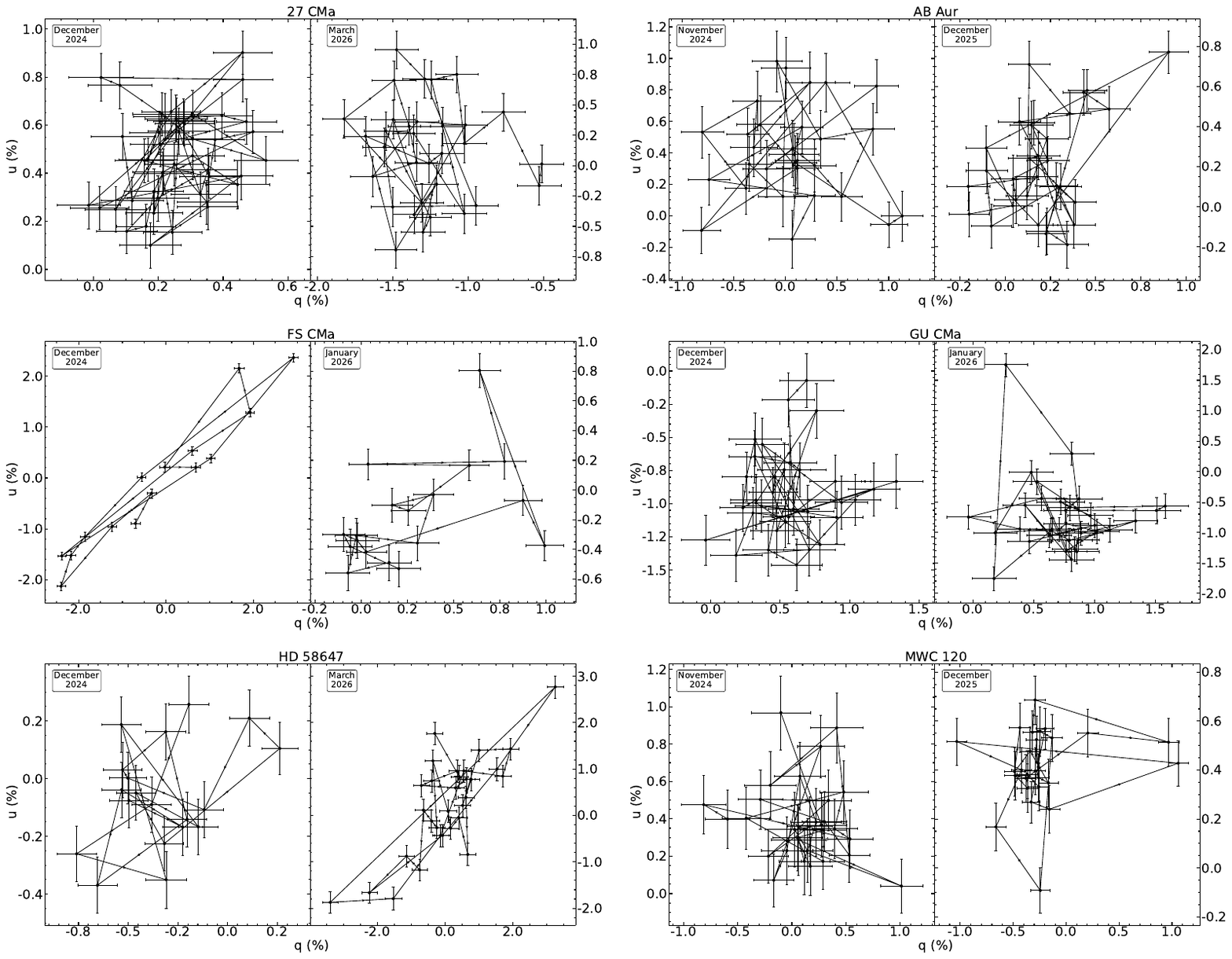}  
  \caption{Multi-epoch qu plots for Herbig Ae/Be stars 27 CMa, AB Aur, FS CMa, GU CMa, HD 58647, and MWC 120. The arrows on the lines indicate increasing wavelength along the spectra.}
  \label{Fig-Herbig_qu_1}
\end{figure*}
%%%%%%%%%%%%%%%%%%%%%%%%%%%%%%%%%%%%%%%%%%%%%%%%%%

%%%%%%%%%%%%%%%%%%%%%%%%%%%%%%%%%%%%%%%%%%%%%%%%%
\begin{figure*}
  \centering
  \includegraphics[angle=90, width=\textwidth]{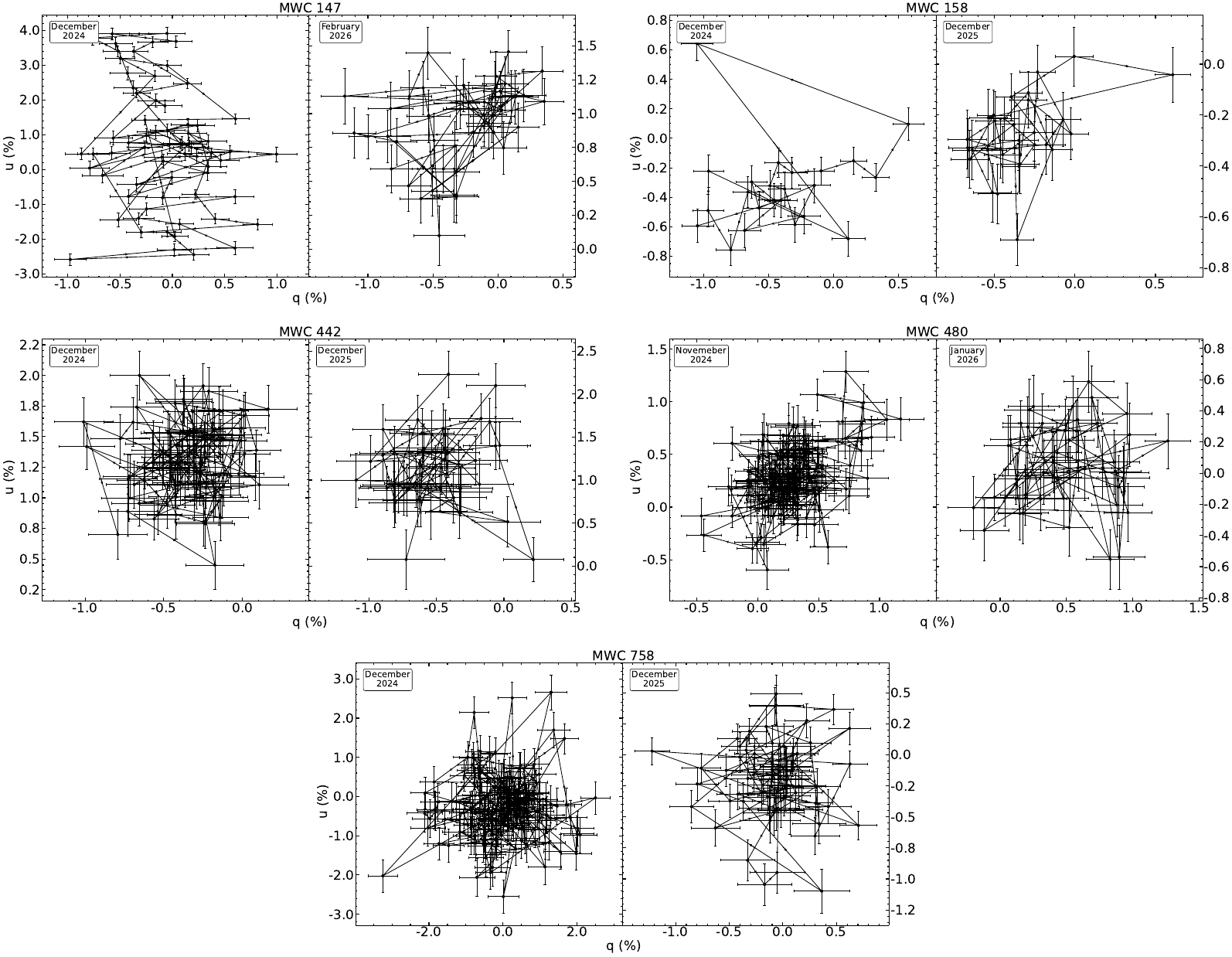}  
  \caption{Same as \ref{Fig-Herbig_qu_1} but for Herbig Ae/Be stars MWC 147, MWC 158, MWC 442, MWC 480, and MWC 758.}
  \label{Fig-Herbig_qu_2}
\end{figure*}
%%%%%%%%%%%%%%%%%%%%%%%%%%%%%%%%%%%%%%%%%%%%%%%%%%

%%%%%%%%%%%%%%%%%%%%%%%%%%%%%%%%%%%%%%%%%%%%%%%%%
\begin{figure*}
  \centering
  \includegraphics[angle=90, width=\textwidth]{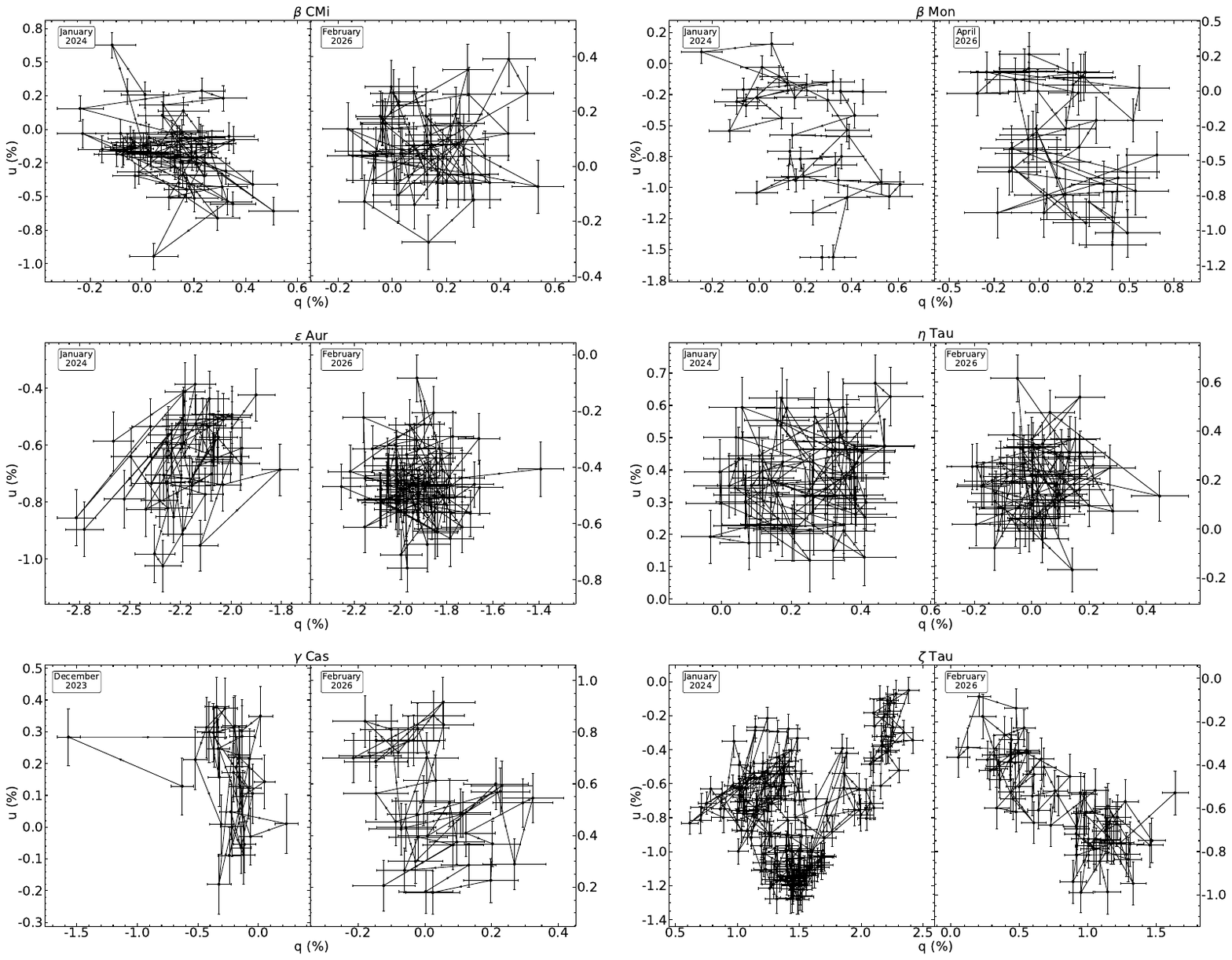}  
  \caption{Same as \ref{Fig-Herbig_qu_1} but for classical Be stars $\beta$ CMi, $\beta$ Mon, $\epsilon$ Aur, $\eta$ Tau, $\gamma$ Cas, and $\zeta$ Tau.}
  \label{Fig-Be_qu_1}
\end{figure*}
%%%%%%%%%%%%%%%%%%%%%%%%%%%%%%%%%%%%%%%%%%%%%%%%%%

%%%%%%%%%%%%%%%%%%%%%%%%%%%%%%%%%%%%%%%%%%%%%%%%%
\begin{figure*}
  \centering
  \includegraphics[angle=90, width=0.65\textwidth]{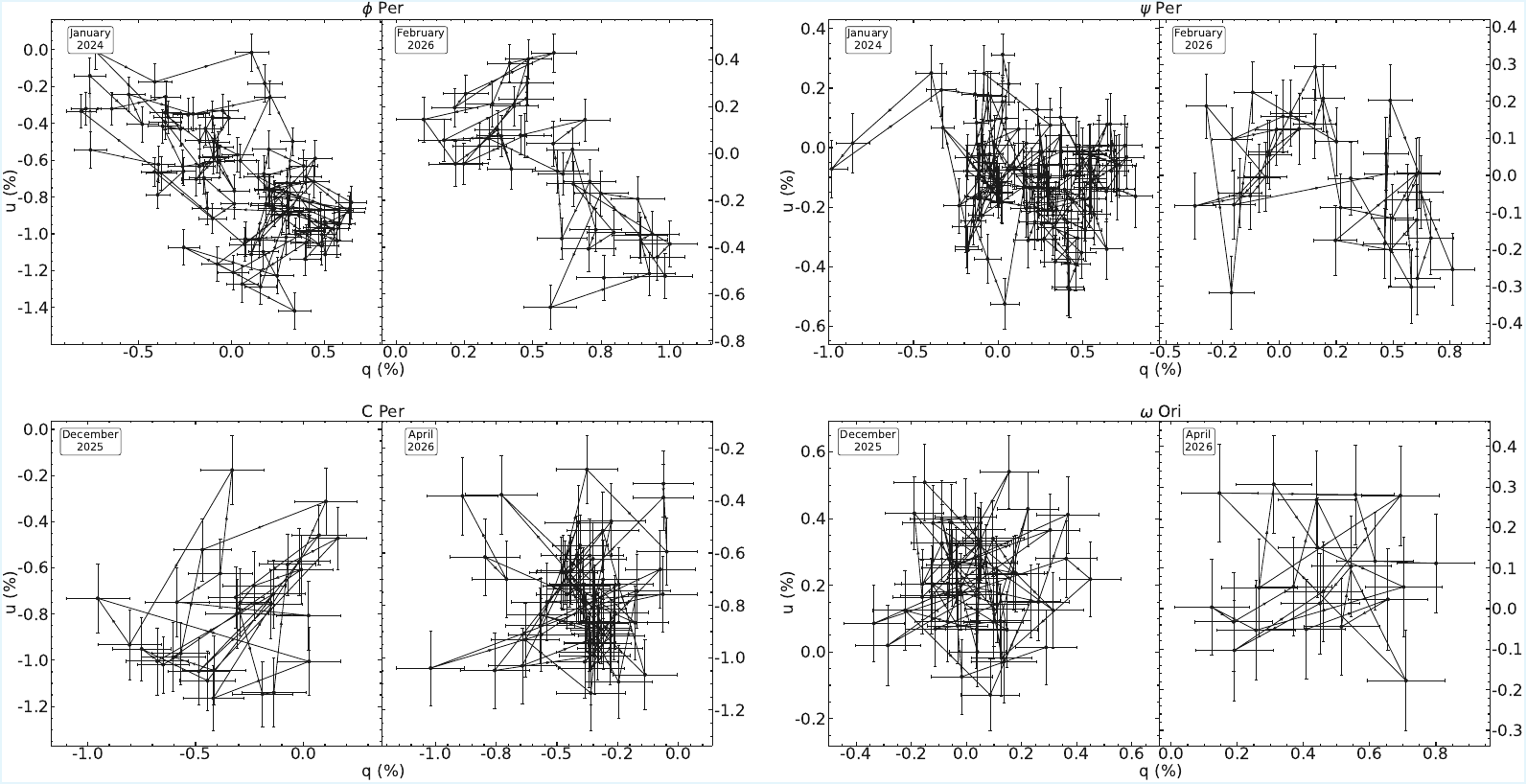}  
  \caption{Same as \ref{Fig-Herbig_qu_1} but for classical Be stars $\phi$ Per, $\psi$ Per, C Per, and $\omega$ Ori.}
  \label{Fig-Be_qu_2}
\end{figure*}
%%%%%%%%%%%%%%%%%%%%%%%%%%%%%%%%%%%%%%%%%%%%%%%%%%

%% For this sample we use BibTeX plus aasjournalv7.bst to generate the

\bibliography{Bibliography}{}

@inproceedings{ikeda2003development,
  title={Development of the high-resolution spectropolarimeter: LIPS},
  author={Ikeda, Yuji and Akitaya, Hiroshi and Matsuda, Kentaro and Kawabata, Koji S and Seki, Munezo and Hirata, Ryuko and Okazaki, Akira},
  booktitle={Polarimetry in Astronomy},
  volume={4843},
  pages={437--447},
  year={2003},
  organization={SPIE}
}

@article{arasaki2015very,
  title={The very precise echelle spectropolarimeter on the Araki telescope (VESPolA)},
  author={Arasaki, Takayuki and Ikeda, Yuji and Shinnaka, Yoshiharu and Itose, Chisato and Nakamichi, Akika and Kawakita, Hideyo},
  journal={Publications of the Astronomical Society of Japan},
  volume={67},
  number={3},
  pages={35},
  year={2015},
  publisher={Oxford University Press}
}

@inproceedings{kumar2022designs,
  title={Designs of Mt. Abu faint object spectrograph and camera-echelle polarimeter (M-FOSC-EP) and its prototype: spectro-polarimeters for PRL 1.2 m and 2.5 m Mt. Abu Telescopes, India},
  author={Kumar, Vipin and Srivastava, Mudit K and Dixit, Vaibhav and Mistry, Bhavesh and Lad, Kevikumar and Patel, Ankita and Rajpurohit, Arvind S},
  booktitle={Ground-based and Airborne Instrumentation for Astronomy IX},
  volume={12184},
  pages={1696--1714},
  year={2022},
  organization={SPIE}
}

@inproceedings{srivastava2024development,
  title={Development of ProtoPol: a medium resolution echelle spectro-polarimeter for PRL 1.2 m and 2.5 m telescopes, Mt Abu, India},
  author={Srivastava, Mudit K and Maiti, Arijit and Kumar, Vipin and Mistry, Bhaveshkumar and Patel, Ankita and Dixit, Vaibhav and Lad, Kevikumar},
  booktitle={Ground-based and Airborne Instrumentation for Astronomy X},
  volume={13096},
  pages={577--592},
  year={2024},
  organization={SPIE}
}

@inproceedings{donati2003espadons,
  title={ESPaDOnS: An Echelle SpectroPolarimetric device for the observation of stars at CFHT},
  author={Donati, J-F},
  booktitle={Solar Polarization},
  volume={307},
  pages={41},
  year={2003}
}

@article{patat2006error,
  title={Error Analysis for Dual-Beam Optical Linear Polarimetry1},
  author={Patat, Ferdinando and Romaniello, Martino},
  journal={Publications of the Astronomical Society of the Pacific},
  volume={118},
  number={839},
  pages={146},
  year={2006},
  publisher={IOP Publishing}
}

@article{oudmaijer1999halpha,
  title={H$\alpha$ spectropolarimetry of B [e] and Herbig Be stars},
  author={Oudmaijer, Ren{\'e} D and Drew, Janet E},
  journal={Monthly Notices of the Royal Astronomical Society},
  volume={305},
  number={1},
  pages={166--180},
  year={1999},
  publisher={Blackwell Science Ltd 23 Ainslie Place, Edinburgh EH3 6AJ, UK. Telephone~…}
}

@article{vink2002probing,
  title={Probing the circumstellar structure of Herbig Ae/Be stars},
  author={Vink, Jorick S and Drew, Janet E and Harries, Tim J and Oudmaijer, Ren{\'e} D},
  journal={Monthly Notices of the Royal Astronomical Society},
  volume={337},
  number={1},
  pages={356--368},
  year={2002},
  publisher={The Royal Astronomical Society}
}

@article{alecian2013high,
  title={A high-resolution spectropolarimetric survey of herbig ae/be stars--i. observations and measurements},
  author={Alecian, E and Wade, GA and Catala, C and Grunhut, JH and Landstreet, JD and Bagnulo, S and B{\"o}hm, T and Folsom, CP and Marsden, S and Waite, I},
  journal={Monthly Notices of the Royal Astronomical Society},
  volume={429},
  number={2},
  pages={1001--1026},
  year={2013},
  publisher={The Royal Astronomical Society}
}

@article{de1994new,
  title={A new catalogue of members and candidate members of the Herbig Ae/Be (HAEBE) stellar group},
  author={De Winter, D and Perez, MR and others},
  journal={Astronomy and Astrophysics Suppl., Vol. 104, p. 315-339 (1994)},
  volume={104},
  pages={315--339},
  year={1994}
}

@article{vink2005probing,
  title={Probing the circumstellar structures of T Tauri stars and their relationship to those of Herbig stars},
  author={Vink, Jorick S and Drew, Janet E and Harries, Tim J and Oudmaijer, Ren{\'e} D and Unruh, Yvonne},
  journal={Monthly Notices of the Royal Astronomical Society},
  volume={359},
  number={3},
  pages={1049--1064},
  year={2005},
  publisher={Blackwell Science Ltd Oxford, UK}
}

@article{harrington2009spectropolarimetric,
  title={Spectropolarimetric observations of Herbig Ae/Be stars. II. Comparison of spectropolarimetric surveys: Haebe, Be and other emission-line stars},
  author={Harrington, DM and Kuhn, Jeffrey R},
  journal={The Astrophysical Journal Supplement Series},
  volume={180},
  number={1},
  pages={138--181},
  year={2009},
  publisher={The American Astronomical Society}
}

@phdthesis{ababakr2016linear,
  title={Linear Spectropolarimetry Of Herbig Ae/Be Stars},
  author={Ababakr, Karim Mahmood},
  year={2016},
  school={University of Leeds}
}

@article{hillenbrand1992herbig,
  title={Herbig Ae/Be stars-Intermediate-mass stars surrounded by massive circumstellar accretion disks},
  author={Hillenbrand, Lynne A and Strom, Stephen E and Vrba, Frederick J and Keene, Jocelyn},
  journal={Astrophysical Journal, Part 1 (ISSN 0004-637X), vol. 397, no. 2, p. 613-643.},
  volume={397},
  pages={613--643},
  year={1992}
}

@article{bjorkman1998first,
  title={The first ultraviolet and optical spectropolarimetry of the B [e] star HD 50138},
  author={Bjorkman, KS and Miroshnichenko, AS and Bjorkman, JE and Meade, MR and Babler, BL and Code, AD and Anderson, CM and Fox, GK and Johnson, JJ and Weitenbeck, AJ and others},
  journal={The Astrophysical Journal},
  volume={509},
  number={2},
  pages={904},
  year={1998},
  publisher={IOP Publishing}
}

@article{pogodin1997circumstellar,
  title={Circumstellar peculiarities in the unusual Be star HD 50138.},
  author={Pogodin, MA},
  journal={Astronomy and Astrophysics, v. 317, p. 185-192},
  volume={317},
  pages={185--192},
  year={1997}
}

@article{baines2006binarity,
  title={On the binarity of Herbig Ae/Be stars},
  author={Baines, Deborah and Oudmaijer, Ren{\'e} D and Porter, John M and Pozzo, Monica},
  journal={Monthly Notices of the Royal Astronomical Society},
  volume={367},
  number={2},
  pages={737--753},
  year={2006},
  publisher={Blackwell Science Ltd Oxford, UK}
}

@article{mottram2007difference,
  title={On the difference between Herbig Ae and Herbig Be stars},
  author={Mottram, Joseph C and Vink, JS and Oudmaijer, RD and Patel, M},
  journal={Monthly Notices of the Royal Astronomical Society},
  volume={377},
  number={3},
  pages={1363--1374},
  year={2007},
  publisher={Blackwell Publishing Ltd Oxford, UK}
}

@article{beskrovnaya2004active,
  title={Active phenomena in the circumstellar environment of the Herbig Ae star HD 31648},
  author={Beskrovnaya, NG and Pogodin, MA},
  journal={Astronomy \& Astrophysics},
  volume={414},
  number={3},
  pages={955--967},
  year={2004},
  publisher={EDP Sciences}
}

@article{beskrovnaya1999spectroscopic,
  title={Spectroscopic, photometric, and polarimetric study of the Herbig Ae candidate HD 36112},
  author={Beskrovnaya, NG and Pogodin, MA and Miroshnichenko, AS and Savanov, IS and Shakhovskoy, NM and Rostopchina, AN and Kozlova, OV and Kuratov, KS and others},
  journal={Astronomy and Astrophysics, v. 343, p. 163-174 (1999)},
  volume={343},
  pages={163--174},
  year={1999}
}

@article{quirrenbach1997constraints,
  title={Constraints on the geometry of circumstellar envelopes: optical interferometric and spectropolarimetric observations of seven Be stars},
  author={Quirrenbach, A and Bjorkman, KS and Bjorkman, JE and Hummel, CA c-a and Buscher, DF and Armstrong, JT and Mozurkewich, D and Elias Ii, NM and Babler, BL},
  journal={The Astrophysical Journal},
  volume={479},
  number={1},
  pages={477},
  year={1997},
  publisher={IOP Publishing}
}

@article{ikeda2004polarized,
  title={Polarized H$\alpha$ Wings in the Symbiotic Stars AG Draconis and Z Andromedae},
  author={Ikeda, Yuji and Akitaya, Hiroshi and Matsuda, Kentaro and Homma, Ken’ichi and Seki, Munezo and Kawabata, Koji S and Hirata, Ryuko and Okazaki, Akira},
  journal={The Astrophysical Journal},
  volume={604},
  number={1},
  pages={357},
  year={2004},
  publisher={IOP Publishing}
}

@article{schmid1994raman,
  title={Raman scattered emission lines in symbiotic stars: a spectropolarimetric survey},
  author={Schmid, HM and Schild, H},
  journal={Astronomy and Astrophysics (ISSN 0004-6361), vol. 281, no. 1, p. 145-160},
  volume={281},
  pages={145--160},
  year={1994}
}

@article{serkowski1974many,
  title={For many astronomical objects the observed polarization is very small, making high polarimetric accuracy essential. Polarimetric precision can},
  author={Serkowski, K},
  journal={Planets, Stars and Nebulae: Studied with Photopolarimetry},
  volume={23},
  pages={135},
  year={1974},
  publisher={University of Arizona Press}
}

@article{clarke1974observations,
  title={Observations of the Linear Polarization in the H $\beta$ Emission Feature of $\gamma$ Cas},
  author={Clarke, D and McLean, IS and Sweet, PA},
  journal={Monthly Notices of the Royal Astronomical Society},
  volume={167},
  number={1},
  pages={27P--30P},
  year={1974},
  publisher={Oxford University Press Oxford, UK}
}

@article{zickgraf1989polarization,
  title={Polarization characteristics of galactic B (e) stars},
  author={Zickgraf, F-J and Schulte-Ladbeck, RE},
  journal={Astronomy and Astrophysics (ISSN 0004-6361), vol. 214, no. 1-2, April 1989, p. 274-284.},
  volume={214},
  pages={274--284},
  year={1989}
}

@article{schulte1994axisymmetric,
  title={The axisymmetric stellar wind of AG Carinae},
  author={Schulte-Ladbeck, Regina E and Clayton, Geoffrey C and Hillier, D John and Harries, Tim J and Howarth, Ian D},
  journal={The Astrophysical Journal, vol. 429, no. 2, pt. 1, p. 846-856},
  volume={429},
  pages={846--856},
  year={1994}
}

@article{cropper1988spectropolarimetry,
  title={Spectropolarimetry of SN 1987A: observations up to 1987 July 8},
  author={Cropper, Mark and Bailey, Jeremy and McCowage, J and Cannon, RD and Couch, Warrick J and Walsh, JR and Strade, JO and Freeman, F},
  journal={Monthly Notices of the Royal Astronomical Society},
  volume={231},
  number={3},
  pages={695--722},
  year={1988},
  publisher={The Royal Astronomical Society}
}

@article{bjorkman1994spectropolarimetry,
  title={Spectropolarimetry of Nova Cygni 1992: Evidence for an asymmetric geometry},
  author={Bjorkman, KS and Johansen, KA and Nordsieck, KH and Gallagher, JS and Barger, AJ},
  journal={Astrophysical Journal, Part 1 (ISSN 0004-637X), vol. 425, no. 1, p. 247-251},
  volume={425},
  pages={247--251},
  year={1994}
}

@article{behrend2001formation,
  title={Formation of massive stars by growing accretion rate},
  author={Behrend, R and Maeder, A},
  journal={Astronomy \& Astrophysics},
  volume={373},
  number={1},
  pages={190--198},
  year={2001},
  publisher={EDP Sciences}
}

@article{bonnell1998formation,
  title={On the formation of massive stars},
  author={Bonnell, Ian A and Bate, Matthew R and Zinnecker, Hans},
  journal={Monthly Notices of the Royal Astronomical Society},
  volume={298},
  number={1},
  pages={93--102},
  year={1998},
  publisher={Blackwell Science Ltd Oxford, UK}
}

@article{hubrig2006accurate,
  title={Accurate magnetic field measurements of Vega-like stars and Herbig Ae/Be stars},
  author={Hubrig, S and Yudin, RV and Sch{\"o}ller, M and Pogodin, MA},
  journal={Astronomy \& Astrophysics},
  volume={446},
  number={3},
  pages={1089--1094},
  year={2006},
  publisher={EDP Sciences}
}

@article{mclean1979interpretation,
  title={Interpretation of the intrinsic polarizations of early-type emission-line stars},
  author={McLean, Ian S},
  journal={Monthly Notices of the Royal Astronomical Society},
  volume={186},
  number={2},
  pages={265--285},
  year={1979},
  publisher={The Royal Astronomical Society}
}

@article{balona1991appearance,
  title={Appearance of Beta Cephei pulsations in the Be star 27 CMa},
  author={Balona, Luis A and Rozowsky, Joel},
  journal={Monthly Notices of the Royal Astronomical Society},
  volume={251},
  number={1},
  pages={66P--68P},
  year={1991},
  publisher={The Royal Astronomical Society}
}

@article{labadie2022classifying,
  title={Classifying Be star variability with TESS. I. The southern ecliptic},
  author={Labadie-Bartz, Jonathan and Carciofi, Alex C and Henrique de Amorim, Tajan and Rubio, Amanda and Luiz Figueiredo, Andr{\'e} and Ticiani dos Santos, Pedro and Thomson-Paressant, Keegan},
  journal={The Astronomical Journal},
  volume={163},
  number={5},
  pages={226},
  year={2022},
  publisher={The American Astronomical Society}
}

@article{oudmaijer2001export,
  title={EXPORT: Optical photometry and polarimetry of Vega-type and pre-main sequence stars},
  author={Oudmaijer, RD and Palacios, J and Eiroa, C and Davies, JK and De Winter, D and Ferlet, R and Garz{\'o}n, F and Grady, CA and Cameron, A and Deeg, HJ and others},
  journal={Astronomy \& Astrophysics},
  volume={379},
  number={2},
  pages={564--578},
  year={2001},
  publisher={EDP Sciences}
}

@article{beskrovnaya1999ma,
  title={MA Pogodin, AS Miroshnichenko, PS Th6, et al},
  author={Beskrovnaya, NG},
  journal={Astron. Astrophys},
  volume={343},
  pages={163},
  year={1999}
}

@article{grady2005coronagraphic,
  title={Coronagraphic imaging of pre-main-sequence stars with the Hubble space telescope space telescope imaging spectrograph. I. The Herbig Ae stars},
  author={Grady, CA and Woodgate, BE and Bowers, CW and Gull, TR and Sitko, ML and Carpenter, WJ and Lynch, DK and Russell, RW and Perry, RB and Williger, GM and others},
  journal={The Astrophysical Journal},
  volume={630},
  number={2},
  pages={958--975},
  year={2005}
}

@article{mannings1997rotating,
  title={A rotating disk of gas and dust around a young counterpart to $\beta$ Pictoris},
  author={Mannings, Vincent and Koerner, David W and Sargent, Anneila I},
  journal={Nature},
  volume={388},
  number={6642},
  pages={555--557},
  year={1997},
  publisher={Nature Publishing Group UK London}
}

@article{vrba1979observations,
  title={Observations and evaluation of the polarization in Herbig Ae/Be stars},
  author={Vrba, FJ and Schmidt, GD and Hintzen, PM},
  journal={Astrophysical Journal, Part 1, vol. 227, Jan. 1, 1979, p. 185-196.},
  volume={227},
  pages={185--196},
  year={1979}
}

@article{poeckert1977linear,
  title={Linear polarization of H-alpha in the Be star gamma Cassiopeiae},
  author={Poeckert, R and Marlborough, JM},
  journal={Astrophysical Journal, Part 1, vol. 218, Nov. 15, 1977, p. 220-226. Research supported by the National Research Council of Canada.},
  volume={218},
  pages={220--226},
  year={1977}
}

@article{mclean1978polarisation,
  title={Polarisation by Thomson scattering in optically thin stellar envelopes. III. A statistical study of the oblateness and rotation of Be star envelopes.},
  author={McLean, IS and Brown, JC},
  journal={Astronomy and Astrophysics, Vol. 69, p. 291-296},
  volume={69},
  pages={291--296},
  year={1978}
}

@article{quirrenbach1993asymmetric,
  title={The asymmetric envelope of gamma cassiopeiae observed with the MK III optical interferometer},
  author={Quirrenbach, A and Hummel, CA and Buscher, DF and Armstrong, JT and Mozurkewich, D and Elias, NM},
  journal={Astrophysical Journal Letters v. 416, p. L25},
  volume={416},
  pages={L25},
  year={1993}
}

@article{quirrenbach1994maximum,
  title={Maximum-entropy maps of the Be shell star zeta Tauri from optical long-baseline interferometry},
  author={Quirrenbach, A and Buscher, DF and Mozurkewich, D and Hummel, CA and Armstrong, JT},
  journal={Astronomy \& Astrophysics},
  volume={283},
  pages={L13},
  year={1994}
}

@article{herbig1960spectra,
  title={The spectra of Be-and Ae-type stars associated with nebulosity},
  author={Herbig, George H},
  journal={Astrophysical Journal Supplement, vol. 4, p. 337},
  volume={4},
  pages={337},
  year={1960}
}

@article{slettebak1988stars,
  title={The be stars},
  author={Slettebak, Arne},
  journal={Publications of the Astronomical Society of the Pacific},
  volume={100},
  number={629},
  pages={770--784},
  year={1988},
  publisher={The Astronomical Society of the Pacific}
}

@article{rivinius2013classical,
  title={Classical Be stars: Rapidly rotating B stars with viscous Keplerian decretion disks},
  author={Rivinius, Thomas and Carciofi, Alex C and Martayan, Christophe},
  journal={The Astronomy and Astrophysics Review},
  volume={21},
  number={1},
  pages={69},
  year={2013},
  publisher={Springer}
}

@article{brown1977polarisation,
  title={Polarisation by Thomson scattering in optically thin stellar envelopes. I. Source star at centre of axisymmetric envelope},
  author={Brown, John C and McLean, Ian S},
  journal={Astronomy and Astrophysics, Vol. 57, p. 141 (1977)},
  volume={57},
  pages={141},
  year={1977}
}

@article{fox1991stellar,
  title={Stellar occultation of polarized light from circumstellar electrons. III-General axisymmetric envelopes},
  author={Fox, Geoffrey K},
  journal={Astrophysical Journal, Part 1 (ISSN 0004-637X), vol. 379, Oct. 1, 1991, p. 663-675. Research supported by Department of Health and Social Security of England.},
  volume={379},
  pages={663--675},
  year={1991}
}

@article{wood1996effect,
  title={The effect of multiple scattering on the polarization from axisymmetric circumstellar envelopes. I. Pure Thomson scattering envelopes},
  author={Wood, Kenneth and Bjorkman, JE and Whitney, Barbara A and Code, Arthur D},
  journal={Astrophysical Journal v. 461, p. 828},
  volume={461},
  pages={828},
  year={1996}
}

@article{bjorkman1994effects,
  title={The effects of gravity darkening on the ultraviolet continuum polarization produced by circumstellar disks},
  author={Bjorkman, JE and Bjorkman, KS},
  journal={Astrophysical Journal, Part 1 (ISSN 0004-637X), vol. 436, no. 2, p. 818-830},
  volume={436},
  pages={818--830},
  year={1994}
}

@article{BrittainHerbig,
  title={Herbig Stars: A Quarter Century of Progress},
  author={Brittain, Sean D. and Inga Kamp and Gwendolyn Meeus and René D. Oudmaijer and L. B. F. M. Waters},
  journal={Space Science Reviews v. 219, no. 7},
  volume={219},
  number={7},
  year={2023}
}

@article{kuzmychov2017first,
  title={First spectropolarimetric measurement of a brown dwarf magnetic field in molecular bands},
  author={Kuzmychov, Oleksii and Berdyugina, Svetlana V and Harrington, David M},
  journal={The Astrophysical Journal},
  volume={847},
  number={1},
  pages={60},
  year={2017},
  publisher={The American Astronomical Society}
}

@article{phan2009magnetic,
  title={Magnetic field topology in low-mass stars: spectropolarimetric observations of M dwarfs},
  author={Phan-Bao, Ngoc and Lim, Jeremy and Donati, Jean-Francois and Johns-Krull, Christopher M and Martin, Eduardo L},
  journal={The Astrophysical Journal},
  volume={704},
  number={2},
  pages={1721--1729},
  year={2009},
  publisher={The American Astronomical Society}
}

@article{fouque2023spirou,
  title={The SPIRou legacy survey-Rotation period of quiet M dwarfs from circular polarization in near-infrared spectral lines: The SPIRou APERO analysis},
  author={Fouqu{\'e}, Pascal and Martioli, Eder and Donati, J-F and Lehmann, Lisa Theres and Zaire, Bonnie and Bellotti, Stefano and Gaidos, Eric and Morin, Julien and Moutou, Claire and Petit, Pascal and others},
  journal={Astronomy \& Astrophysics},
  volume={672},
  pages={A52},
  year={2023},
  publisher={EDP Sciences}
}

@article{donati2020spirou,
  title={SPIRou: NIR velocimetry and spectropolarimetry at the CFHT},
  author={Donati, JF and Kouach, D and Moutou, C and Doyon, R and Delfosse, X and Artigau, E and Baratchart, S and Lacombe, M and Barrick, G and H{\'e}brard, G and others},
  journal={Monthly Notices of the Royal Astronomical Society},
  volume={498},
  number={4},
  pages={5684--5703},
  year={2020},
  publisher={Oxford University Press}
}

@article{kawakita2019high,
  title={High-resolution Optical Spectropolarimetry of Nova V339 Del: Spatial Distribution of Nova Ejecta during the Early Phase of Explosion},
  author={Kawakita, H and Shinnaka, Y and Arai, A and Arasaki, T and Ikeda, Y},
  journal={The Astrophysical Journal},
  volume={872},
  number={2},
  pages={120},
  year={2019},
  publisher={The American Astronomical Society}
}

@article{srivastava2026development,
  title={Development of ProtoPol: a medium resolution echelle spectro-polarimeter for PRL telescopes, Mt. Abu, India—Part I: the design, development, and laboratory characterization},
  author={Srivastava, Mudit K and Maiti, Arijit and Kumar, Vipin and Mistry, Bhaveshkumar and Patel, Ankita and Dixit, Vaibhav and Lad, Kevikumar A},
  journal={Journal of Astronomical Telescopes, Instruments, and Systems},
  volume={12},
  number={2},
  pages={028001--028001},
  year={2026},
  publisher={Society of Photo-Optical Instrumentation Engineers}
}

@article{maiti2026development,
  title={Development of ProtoPol: a medium-resolution echelle spectro-polarimeter for PRL telescopes, Mt. Abu, India—Part II: the data-reduction pipeline, on-sky characterization and performance verification, and first science results},
  author={Maiti, Arijit and Srivastava, Mudit K and Kumar, Vipin and Mistry, Bhaveshkumar and Patel, Ankita and Dixit, Vaibhav and Pandey, Ruchi and Chitroda, Jay},
  journal={Journal of Astronomical Telescopes, Instruments, and Systems},
  volume={12},
  number={2},
  pages={028002--028002},
  year={2026},
  publisher={Society of Photo-Optical Instrumentation Engineers}
}

@article{hernandez2004spectral,
  title={Spectral analysis and classification of Herbig Ae/Be stars},
  author={Hern{\'a}ndez, Jes{\'u}s and Calvet, Nuria and Briceno, C{\'e}sar and Hartmann, Lee and Berlind, Perry},
  journal={The Astronomical Journal},
  volume={127},
  number={3},
  pages={1682--1701},
  year={2004}
}

@article{maeder2000evolution,
  title={The evolution of rotating stars},
  author={Maeder, Andr{\'e} and Meynet, Georges},
  journal={Annual Review of Astronomy and Astrophysics},
  volume={38},
  number={1},
  pages={143--190},
  year={2000},
  publisher={Annual Reviews 4139 El Camino Way, PO Box 10139, Palo Alto, CA 94303-0139, USA}
}

@article{ignace1996equatorial,
  title={Equatorial wind compression effects across the HR diagram},
  author={Ignace, R and Cassinelli, JP and Bjorkman, JE},
  journal={Astrophysical Journal v. 459, p. 671},
  volume={459},
  pages={671},
  year={1996}
}

@article{wheelwright2010mass,
  title={The mass ratio and formation mechanisms of Herbig Ae/Be star binary systems},
  author={Wheelwright, HE and Oudmaijer, RD and Goodwin, SP},
  journal={Monthly Notices of the Royal Astronomical Society},
  volume={401},
  number={2},
  pages={1199--1218},
  year={2010},
  publisher={Blackwell Publishing Ltd Oxford, UK}
}

@article{garcia2016investigating,
  title={Investigating the origin and spectroscopic variability of the near-infrared H I lines in the Herbig star VV Ser},
  author={Garcia Lopez, Rebeca and Kurosawa, Ryuichi and Caratti o Garatti, Alessio and Kreplin, Alexander and Weigelt, Gerd and Tambovtseva, Larisa V and Grinin, Vladimir P and Ray, Thomas P},
  journal={Monthly Notices of the Royal Astronomical Society},
  volume={456},
  number={1},
  pages={156--170},
  year={2016},
  publisher={Oxford University Press}
}

@article{maiti2026discovery,
  title={Discovery of Variable Polarization in the H $\alpha$ Profile of the Symbiotic Star Y Gem: A Case for Orbital-phase-dependent Variation in Raman-scattered Ly $\beta$ Emission},
  author={Maiti, Arijit and Srivastava, Mudit K and Kumar, Vipin},
  journal={The Astrophysical Journal},
  volume={1005},
  number={1},
  pages={123},
  year={2026},
  publisher={The American Astronomical Society}
}
\bibliographystyle{aasjournalv7}



\end{document}